\documentclass[
prd 
,showpacs, amssymb, superscriptaddress, aps
]{revtex4}
\usepackage{graphicx}
\input{epsf}

\usepackage{amsmath,amssymb}
\usepackage{bm}
\usepackage{times}

\newcommand{\dalm}{\kern1pt\vbox{\hrule height 0.9pt\hbox{\vrule width 0.9pt
\hskip 2.5pt\vbox{\vskip 5.5pt}\hskip 3pt\vrule width 0.3pt}\hrule height 0.3pt}
\kern1pt}

\newcommand{\be}{\begin{eqnarray}}
\newcommand{\ee}{\end{eqnarray}}
\newcommand{\beq}{\begin{eqnarray}}
\newcommand{\eeq}{\end{eqnarray}}

\begin{document}



\title{Systematical study of pulsar light curves with special relativistic effects}

\author{Hajime Sotani}
\email{sotani@yukawa.kyoto-u.ac.jp}
\affiliation{Division of Theoretical Astronomy, National Astronomical Observatory of Japan, 2-21-1 Osawa, Mitaka, Tokyo 181-8588, Japan}

\author{Umpei Miyamoto}
\affiliation{Research and Education Center for Comprehensive Science, Akita Prefectural University, Akita 015-0055, Japan}

\date{\today}

\begin{abstract}
We systematically study pulsar light curves, taking into account the special relativistic effect, i.e., the Doppler factor due to the fast spin of the neutron stars, together with the time delay, which comes from the difference of the travel times depending on the position of the spots. For this purpose, first we derive the basic equations with the general expression of the metric for the static, spherically symmetric spacetime, where for simplicity we adopt the pointlike spot approximation for the antipodal spots associated with the magnetic polar cap model. Then, we calculate the light curves from the neutron star models in general relativity, with various angle between rotational and magnetic axes and the inclination angle. As the results, unlike the case for a slowly rotating stellar model, we find that the light curve from a fast rotating stellar model depends not only the stellar compactness but also the stellar radius. We also find that the amplitude of the light curve becomes larger as the stellar radius increases and as the stellar compactness decreases. Thus, via careful observations of the light curves from the rotating neutron star, one would determine the stellar compactness together with the stellar radius, if it rotates fast enough. 
\end{abstract}

%

\maketitle

\section{Introduction}
\label{sec:I}

Neutron stars, which are produced by supernovae, are one of the most suitable candidates for testing the fundamental physics. In fact, the density inside the star significantly exceeds the standard nuclear density, and the magnetic and gravitational fields inside/around the star become very strong \cite{shapiro-teukolsky}. Due to such a high density inside the star, the equation of state (EOS) for neutron-star matter cannot be constrained only with terrestrial nuclear experiments. Therefore, direct observations of neutron stars help to constrain the EOS. In this context, the discoveries of the $2M_\odot$ neutron stars \cite{D2010,A2013} are very important, by which some of soft EOSs have been ruled out. The strong gravitational field is possible to observationally verify the theory of gravity. Actually, there are many tests of general relativity in the weak field regime, but the tests in the strong field regime are still very poor. That is, the gravitational theory in the strong field regime may deviate from general relativity. If so, one can test the gravitational theory via the observations of neutron stars (e.g., \cite{Berti2015,SK2004,Sotani2014,Sotani2014a}).

In order to see properties of the strong gravitational field, the light bending is also one of the important phenomena. Unlike the Newtonian theory, the photon path is bent due to the relativistic effect. As a result, the photon radiating from the backside of a neutron star may be observed \cite{PFC1983}. This is a phenomenon similar to the strong lensing effect around a black hole (e.g., \cite{VE2000,Bozza2002,SM2015}). Thus, to consider the light bending by a neutron star, the stellar compactness, i.e., the ratio of the mass to the radius, is an important property, because it is a kind of parameter expressing how strong the gravitational field around/inside the star becomes. So, by observing the light curve from the rotating neutron star, one may determine the stellar compactness, which helps us to constrain the EOS for neutron-star matter \cite{POC2014,Bogdanov2016}. In fact, such an attempt could come true soon by the operating Neutron star Interior Composition ExploreR mission \cite{NICER}. A light curve can be calculated by numerical integration if one chooses the angle between the magnetic and rotational axes and the inclination angle, while several approximation relations have been proposed to easily calculate the pulse profiles \cite{LL95,Beloborodov2002,PG03,PB06}. These approximations may be useful for neutron stars with small compactness \cite{SM2017}. On the other hand, since the light curve should depend on the spacetime geometry outside the star, one may test the gravitational geometry via the observations of the pulse profiles from the rotating neutron star \cite{SM2017,S2017,SY2018}.

Most of the previous studies have been done in the Schwarzschild spacetime, where the rotational effects are neglected. However, if one considers the pulse profiles from a fast rotating neutron star, the rotational effects should be taken into account \cite{PG03,PB06,CLM05,PO2014}. In fact, the fastest rotational frequency of pulsar discovered up to now is 716 Hz \cite{716Hz}, where the light curves must be different from the expectation obtained in the Schwarzschild spacetime. In any way, since so far the systematical studies for the light curves with the rotational effects are very few, in this paper we will examine the pulse profiles with various stellar models and see the dependence on the stellar properties. We adopt the geometric units, $c=G=1$, where $c$ and $G$ denote the speed of light and the gravitational constant, respectively, and the metric signature is $(-,+,+,+)$.

\section{Photon radiating from hot spots}
\label{sec:II}

\begin{figure}
\begin{center}
\includegraphics[scale=0.4]{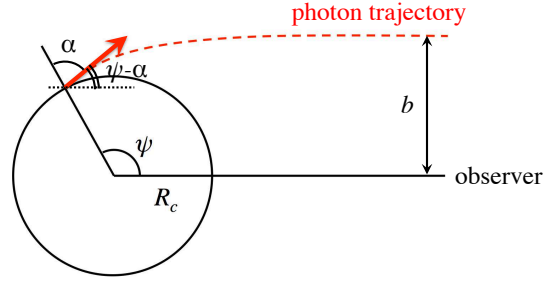} 
\end{center}
\caption{
Image of the photon trajectory from the stellar surface at the position angle of $\psi$ with the dash line.  $R_c$, $b$, and $\alpha$ respectively denote the stellar radius, the impact parameter given by Eq. (\ref{eq:b}), and the emission angle, where $\psi-\alpha$ corresponds to the bending angle.
}
\label{fig:trajectory}
\end{figure}

The metric for the static, spherically symmetric spacetime is generally expressed as
\begin{align}
  ds^2 &= g_{\mu\nu}dx^\mu dx^\nu \nonumber \\
          &= -A(r)dt^2 + B(r)dr^2 + C(r)\left(d\theta^2 + \sin^2\theta d\psi^2\right),
\end{align}
where we especially focus on the asymptotically flat spacetime, i.e., $A(r)\to 1$, $B(r)\to 1$, and $C(r)\to r^2$ as $r\to \infty$. We remark that the circumference radius, $r_c$, is associated with the radial coordinate, $r$, via the relation of $r_c^2=C(r)$. Due to the nature of spherical symmetry, one can assume without loss of generality that the photon trajectory is in the plane with $\theta=\pi/2$. Then, choosing the direction of the observer far from the central object as $\psi=0$, the angle of the hot-spot position at the stellar surface, $r=R$, is given by
\begin{equation}
  \psi(R) = \int_R^\infty \frac{dr}{C}\left[\frac{1}{AB}\left(\frac{1}{b^2}-\frac{A}{C}\right)\right]^{-1/2},   \label{eq:psi}
\end{equation}
where $b$ is an impact parameter given by 
\begin{equation}
 b = \sin\alpha \sqrt{\frac{C(R)}{A(R)}}.  \label{eq:b}
\end{equation}
Here, $\alpha$ denotes the emission angle, which is the angle between the direction of photon radiation and the normal vector at the hot spot, as shown in Fig. \ref{fig:trajectory}. We remark that the physical stellar radius, $R_c$, should be considered as $R_c^2=C(R)$. By combining Eqs. (\ref{eq:psi}) and (\ref{eq:b}), one can numerically derive the relation between $\psi(R)$ and $\alpha$ for given $R$. We remark that $\psi$ increases as $\alpha$ increases and becomes maximum $\psi_{\rm cri}$ when $\alpha=\pi/2$. In the present study, we simply focus on only the case of $\psi_{\rm cri}<\pi$, i.e., the invisible zone exists \cite{highMR}.

\begin{figure}
\begin{center}
\includegraphics[scale=0.4]{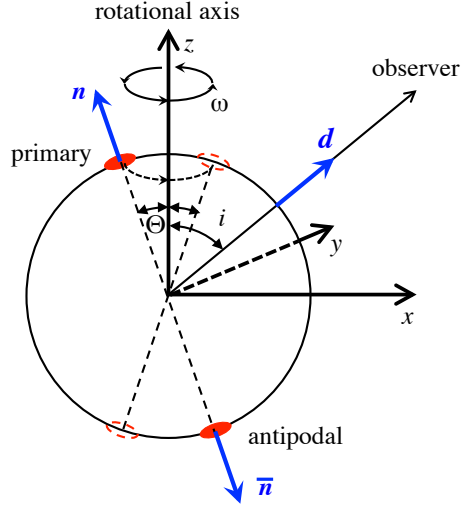} 
\end{center}
\caption{
Image of the hot spots on the rotating star with angular velocity $\omega$, where two hot spots are associated with the magnetic polar caps. 
}
\label{fig:pulsar}
\end{figure}

Now, as shown in Fig. \ref{fig:pulsar}, we consider the hot spots on the rotating neutron star with the angular velocity $\omega$ measured by the observer. In particular, we consider that two hot spots are associated with the magnetic polar caps, where the hot spot closer to the observer is identified with primary while the other is antipodal. The unit vector pointing toward the observer and the normal vector at the primary (antipodal) hot spot are denoted by $\bm{d}$ and $\bm{n}$ ($\bar{\bm{n}}$), respectively. The angle between the rotational axis and $\bm{n}$ is $\Theta$, and the angle between the rotational axis and $\bm{d}$ is $i$, where the angles $\Theta$ and $i$ can be chosen in the range of $\Theta\in[0,\pi/2]$ and $i\in[0,\pi/2]$. Setting the Cartesian coordinate system ($x,y,z$) in such a way that the $z$-axis should be along the rotational axis and $\bm{d}$ should be on the $z$-$x$ plane, and assuming that the primary spot comes closest to the observer at $t=0$, $\bm{d}$ and $\bm{n}$ can be expressed as
\begin{gather}
   \bm{d} = \left[\sin i, 0, \cos i \right], \\
   \bm{n} = \left[\sin\Theta\cos(\omega t), \sin\Theta\sin(\omega t), \cos\Theta\right],
\end{gather}
which leads to
\begin{align}
  \cos\psi &= \bm{d}\cdot\bm{n} \nonumber \\
      & = \sin i \sin\Theta\cos(\omega t) + \cos i \cos\Theta.  \label{eq:cospsi}
\end{align}
We note that $t$ is the time for the observer far from the neutron star and the photon trajectory is on the plane spanned by $\bm{d}$ and $\bm{n}$ in any time. Expressing the unit vector pointing to the initial direction of the photon emitted from the primary hot spot as $\bm{d}_0$, by definition one can get the relation that $\cos\alpha=\bm{d}_0\cdot\bm{n}$ and $\cos(\psi-\alpha)=\bm{d}\cdot\bm{d}_0$. Thus, $\bm{d}_0$ is written with $\bm{d}$ and $\bm{n}$ as
\begin{equation}
  \bm{d}_0 = {\cal A}\bm{d} + {\cal B}\bm{n},
\end{equation}
where
\begin{equation}
   {\cal A} := \frac{\sin\alpha}{\sin\psi} \ \ {\rm and}\ \ {\cal B}:=\frac{\sin(\psi-\alpha)}{\sin\psi}.
\end{equation}
Furthermore, the velocity of the motion of hot spot, $\bm{v}$, is expressed by
\begin{gather}
  \frac{\bm{v}}{|\bm{v}|} = \left[-\sin(\omega t), \cos(\omega t), 0\right], \\
  |\bm{v}| = \sqrt{C(R)}\,\omega_0\sin\Theta = \sqrt{\frac{C(R)}{A(R)}}\,\omega\sin\Theta,
\end{gather}
where $\omega_0=\omega/\sqrt{A(R)}$ is the angular velocity in the vicinity of the stellar surface.

\begin{figure}
\begin{center}
\includegraphics[scale=0.4]{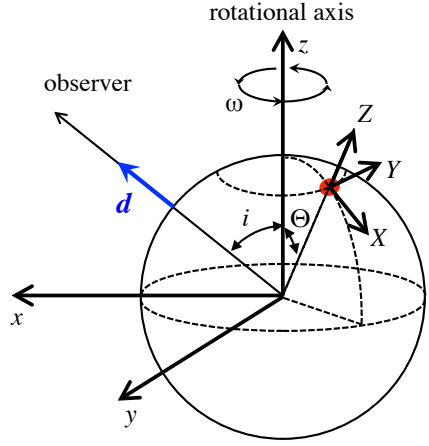} 
\end{center}
\caption{
Relation between the coordinate systems $(x,y,z)$ and $(X,Y,Z)$, where only the primary hot spot is shown.
}
\label{fig:pulsar1}
\end{figure}

At any time, one can choose the instantaneous non-rotating frame $(X,Y,Z)$ on the center of the primary hot spot, where the $Z$-axis coincides with $\bm{n}$, the $Y$-axis is put to the direction of the spot motion, and $X$-axis is along the meridian towards the equator (see Fig. \ref{fig:pulsar1}). In this frame, $\bm{d}_0$ is expressed as
\begin{equation}
  \bm{d}_0 = \left[{\cal A}\left(\sin i\cos\Theta\cos(\omega t) - \cos i\sin\Theta\right), \cos\xi, \cos\alpha \right],
\end{equation}
where $\xi$ denotes the angle between $Y$-axis and $\bm{d}_0$, i.e., 
\begin{equation}
  \cos\xi = \frac{\bm{v}}{|\bm{v}|}\cdot\bm{d}_0 = \frac{{\cal A}\bm{v}}{|\bm{v}|}\cdot\bm{d} = -{\cal A}\sin i\sin(\omega t).
\end{equation}

Now, in order to take into account a special relativistic effect, we consider the Lorentz transformation of the position of hot spot from a non-rotating frame $(X,Y,Z)$ to a co-rotating frame $(X',Y',Z')$, 
\begin{gather}
 {\cal T}' = \gamma({\cal T}-|\bm{v}| Y), \\
 Y' = \gamma(-|\bm{v}| {\cal T} + Y), \\
 X' = X, \\
 Z' = Z,
\end{gather}
where $\gamma=(1-|\bm{v}|^2)^{-1/2}$, while ${\cal T}$ and ${\cal T}'$ are respectively the time in the frame of $(X,Y,Z)$ and $(X',Y',Z')$. Hereafter, the primed quantities denote those measured in the frame corotating with the spot. In the corotating frame, the unit vector, $\bm{d}_0'$, along the photon momentum is given by
\begin{equation}
  \bm{d}_0' = \delta\left[{\cal A}\left(\sin i\cos\Theta\cos(\omega t) - \cos i\sin\Theta\right),\gamma(\cos\xi-|\bm{v}|), \cos\alpha \right],
\end{equation}
where $\delta$ is the Doppler factor given by
\begin{equation}
  \delta = \frac{1}{\gamma(1-|\bm{v}|\cos\xi)}.
\end{equation}
Thus, the emission angle in the corotating frame $\alpha'$, which is the angle between the $Z'$-axis and $\bm{d}_0'$, is given by
\begin{equation}
  \cos\alpha' = \delta\cos\alpha.  \label{eq:cos}
\end{equation}
We also remark that the apparent spot area $dS$ measured in the non-rotating frame is related to the actual spot area $dS'$ measured in the corotating frame as 
\begin{equation}
  dS = \delta dS'
\end{equation}
due to the aberration and special relativistic effect \cite{Ghisellini99}. Finally, one can find that the projected spot area is Lorentz invariant, i.e, $dS\cos\alpha = dS'\cos\alpha'$.

\section{observed flux with time delay}
\label{sec:V}

With the solid angle $d\Omega$ occupied by the spot area $dS'$ on the observer's sky, the observed spectral flux $dF_E$ is given by
\begin{equation}
 dF_E = I_E d\Omega,  \label{dFE}
\end{equation}
where $I_E$ is the specific intensity of radiation with photon energy $E$ for the observer. The solid angle $d\Omega$ is expressed with the impact parameter $b$ as
\begin{equation}
  d\Omega = \frac{b\,db\,d\phi}{D^2}, 
\end{equation}
where $D$ is the distance from the star to the observer and $\phi$ is the azimuthal angle with respect to the direction of rotation around $\bm{d}$. Using Eq. (\ref{eq:b}) and the relations that $dS=C(R)\sin\psi\,d\psi\,d\phi$ and $dS\cos\alpha = dS'\cos\alpha'$, $d\Omega$ is rewritten as
\begin{equation}
  d\Omega = \frac{dS'}{D^2}\frac{\cos\alpha'}{A(R)}\frac{d(\cos\alpha)}{d(\cos\psi)}.  \label{eq:dOme}
\end{equation}

Now, the specific intensity $I_0(E_0,\alpha)$ measured in the vicinity of the stellar surface in non-rotating frame and $I'(E',\alpha')$  in corotating frame, are related by
\begin{gather}
  \frac{I_E}{I_0(E_0,\alpha)} = \left(\frac{E}{E_0}\right)^3 = A(R)^{3/2}, \label{eq:IEI0} \\
  \frac{I_0(E_0,\alpha)}{I'(E',\alpha')} = \left(\frac{E_0}{E'}\right)^3 = \delta^3, \label{eq:I0I'}
\end{gather}
where $E_0$ and $E'$ are respectively the photon energy measured in the vicinity of the stellar surface in non-rotating and in corotating frames \cite{PG03,PB06}. Thus, using Eqs. (\ref{eq:cos}), (\ref{eq:dOme}), (\ref{eq:IEI0}), and (\ref{eq:I0I'}), $dF_E$ given by Eq. (\ref{dFE}) is written as 
\begin{equation}
  dF_E = A(R)^{1/2}\delta^4 I'(E',\alpha')\cos\alpha\frac{d(\cos\alpha)}{d(\cos\psi)}\frac{dS'}{D^2}.
\end{equation}
Thus, the observed bolometric flux $dF$ is
\begin{equation}
  dF := \int_0^\infty (dF_E) dE = A(R)\delta^5 I'(\alpha')\cos\alpha\frac{d(\cos\alpha)}{d(\cos\psi)}\frac{dS'}{D^2}, \label{eq:dF}
\end{equation}
where $I'(\alpha')=\int_0^\infty I'(E',\alpha')dE'$ is the bolometric intensity in corotating frame. In order to derive Eq. (\ref{eq:dF}), we adopt the relation $E/E'=\sqrt{A(R)}\delta$. As in Refs. \cite{Beloborodov2002,PG03,PB06}, by adopting the pointlike spot approximation for simplicity, the observed bolometric flux $F$ from the hot spot is
\begin{equation}
  F := \int dF = A(R)\delta^5 I'(\alpha')\cos\alpha\frac{d(\cos\alpha)}{d(\cos\psi)}\frac{S'}{D^2}, \label{eq:FF}  
\end{equation}
where $S':=\int dS'$ denotes the area of the hot spot. In general $I'(\alpha')$ depends on the emission angle, but as in \cite{Beloborodov2002,SM2017,S2017} we assume the isotropic emission in a local Lorentz frame, i.e., $I'=$const.. Then, the observed bolometric flux is
\begin{equation}
  F(t) = F_1\delta^5 \cos\alpha\frac{d(\cos\alpha)}{d(\cos\psi)}, \ \ F_1:=\frac{I'S'A(R)}{D^2}.  \label{eq:FF1}
\end{equation}
We remark that the limit of $\delta\to 1$ corresponds to the case neglecting the special relativistic effect due to the stellar rotation. 
We also remark that the shape of $F(t)/F_1$ is independent of the choice of $S'$ and $I'$.

For considering the observation, one may have to take into account the time delay $\Delta t$, which comes from the fact that the travel times of radiated photons to the observer depends on the position of the hot spot. From Eq. (\ref{eq:cospsi}), we consider the situation that $\psi$ becomes minimum at $t=0$, where $\alpha$ and $b$ also become minimum. Now, we consider the time delay \cite{TD-PB06}, which is the difference from the travel time when $\psi$ becomes minimum, i.e.,
\begin{equation}
  \Delta t(t) = \int_R^\infty \sqrt{\frac{B}{A}}\left[\left(1-\frac{Ab^2}{C}\right)^{-1/2}-\left(1-\frac{Ab_{\rm min}^2}{C}\right)^{-1/2}\right]dr,
  \label{eq:dt}
\end{equation}
where $b_{\rm min}$ is the impact parameter when $\psi$ becomes minimum, i.e., the minimum impact parameter. With this time delay, the observer time $t_{\rm ob}$ is given by
\begin{equation}
  t_{\rm ob} = t + \Delta t(t). \label{eq:tob}
\end{equation} 
Finally, combining Eqs. (\ref{eq:FF1}) and (\ref{eq:tob}), one can calculate the observed light curve, i.e., $F(t_{\rm {ob}})$. Here, we remark that the classification whether the hot spots are observed or not is the same as in Ref. \cite{SM2017}, which depends on the stellar compactness and spacetime geometry, independently of the introduction of the special relativistic effect. That is, the light bending calculated with Eqs. (\ref{eq:psi}) and (\ref{eq:b}) depends only on the stellar compactness $M/R_c$, while the special relativistic effect depends on not only $M/R_c$ but also $R_c$ through the term of $\sqrt{C(R)}$ in $|\bm{v}|$. Thus, carefully observing the pulse profiles, one may obtain the information about $M/R_c$ and $R_c$.

In order to calculate Eqs. (\ref{eq:psi}) and (\ref{eq:dt}) numerically, we introduce an arbitrary constant $R_*$, where $R_* \gg R$, and assume the asymptotical behaviors of metric functions $A(r)$, $B(r)$, and $C(r)$ as
\begin{gather}
   A(r) = 1 + \frac{a_1}{r} + \frac{a_2}{r^2} + {\cal O}\left(\frac{1}{r^3}\right), \\
   B(r) = 1 + \frac{b_1}{r} + \frac{b_2}{r^2} + {\cal O}\left(\frac{1}{r^3}\right), \\   
   C(r) = r^2\left[1 + \frac{c_1}{r} + \frac{c_2}{r^2} + {\cal O}\left(\frac{1}{r^3}\right)\right],
\end{gather}
where $a_1$, $a_2$, $b_1$, $b_2$, $c_1$, and $c_2$ are some constants depending on the adopted spacetime. 
This is because one can not generally solve the integrations in Eqs. (\ref{eq:psi}) and (\ref{eq:dt}) analytically even with specific metric functions.
Then, Eqs. (\ref{eq:psi}) and (\ref{eq:dt}) can be expressed as
\begin{equation}
  \psi(R) = \int_R^{R_*} \frac{b\sqrt{AB}}{\sqrt{C(C-Ab^2)}}dr 
     + b\left[\frac{1}{R_*} + \frac{a_1 + b_1 - 2c_1}{4R_*^2} + {\cal O}\left(\frac{1}{R_*^3}\right)\right]   \label{eq:psi1}
\end{equation}
and
\begin{equation}
  \Delta t(t) = \int_R^{R_*} \sqrt{\frac{BC}{A}}\left[\frac{1}{\sqrt{C-Ab^2}} - \frac{1}{\sqrt{C-Ab_{\rm min}^2}}\right]dr
      + \left(b^2-b_{\rm min}^2\right)\left[\frac{1}{2R_*} + \frac{a_1 + b_1 - 2c_1}{8R_*^2} + {\cal O}\left(\frac{1}{R_*^3}\right)\right].
  \label{eq:dt1}
\end{equation}
We find that the coefficient of the term of $1/R_*^2$ in Eq. (13) in Ref. \cite{SM2017} is not correct, which should be as in Eq. (\ref{eq:psi1}), although the numerical results in Ref. \cite{SM2017} are almost insensitive to this correction.

\section{Pulse profiles with various stellar models}
\label{sec:IV}

\begin{figure}
\begin{center}
\includegraphics[scale=0.5]{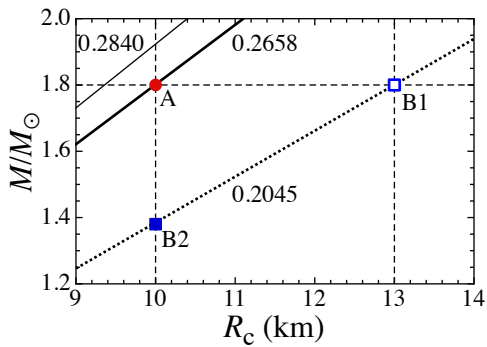} 
\end{center}
\caption{
The stellar models adopted in this study are denoted by A (circle), B1 (open square), and B2 (filled-square). The thin-solid, thick-solid, and dotted lines correspond to the stellar models with $M/R_c=0.2840$, 0.2658, and 0.2045, respectively.  
}
\label{fig:MR}
\end{figure}

In this paper we focus on the case of the Schwarzschild spacetime outside the star, i.e., the metric functions are given by
 \begin{equation}
  A(r) = 1-\frac{2M}{r}, \ \ B(r) = \frac{1}{A(r)}, \ \ C(r) = r^2.
\end{equation}
In order to see the dependence of the pulse profiles on the stellar models, we particularly consider three stellar models as shown in Fig. \ref{fig:MR}, where the circle and open square denote the $1.8M_\odot$ stellar models with $R_c=10$ and 13 km, while the filled square denotes the stellar model with $R_c=10$ km and the same compactness as the model denoted by the open square. Hereafter, these stellar models are referred to as A, B1, and B2. In addition, in Fig. \ref{fig:MR} the thin-solid, thick-solid, and dotted lines denote the stellar models, whose stellar compactness is constant to be $M/R_c=0.2840$, 0.2658, and 0.2045. For the Schwarzschild spacetime outside the star, the thin-solid line corresponds to the critical line at which the invisible zone disappears, i.e., the invisible zone disappears if the star with the compactness larger than 0.2840 \cite{SM2018}. With respect to these stellar models, A, B1, and B2, in order to see the dependence on the stellar rotation, we consider two cases of the rotational frequencies $\nu\equiv \omega/2\pi=0.1$ and 700 Hz. In fact, the fastest rotating pulsar observed up to now is spinning with $\nu=716$ Hz \cite{716Hz}, while the strongly magnetized neutron stars, the so-called magnetars, are slowly spinning with $\nu\simeq 0.1$ Hz. To clarify the rotational frequency in the name of model, for example we refer to the model A with 700 Hz as A(700).
In this study, we adopt the Schwarzschild spacetime outside the star even for considering the fast rotating neutron stars. However, by taking into account the effects of the time delay and Doppler factor, the results even with the Schwarzschild spacetime could be quite similar to the results by the full general relativistic calculations \cite{CLM05}.

\begin{figure}
\begin{center}
\begin{tabular}{cc}
\includegraphics[scale=0.5]{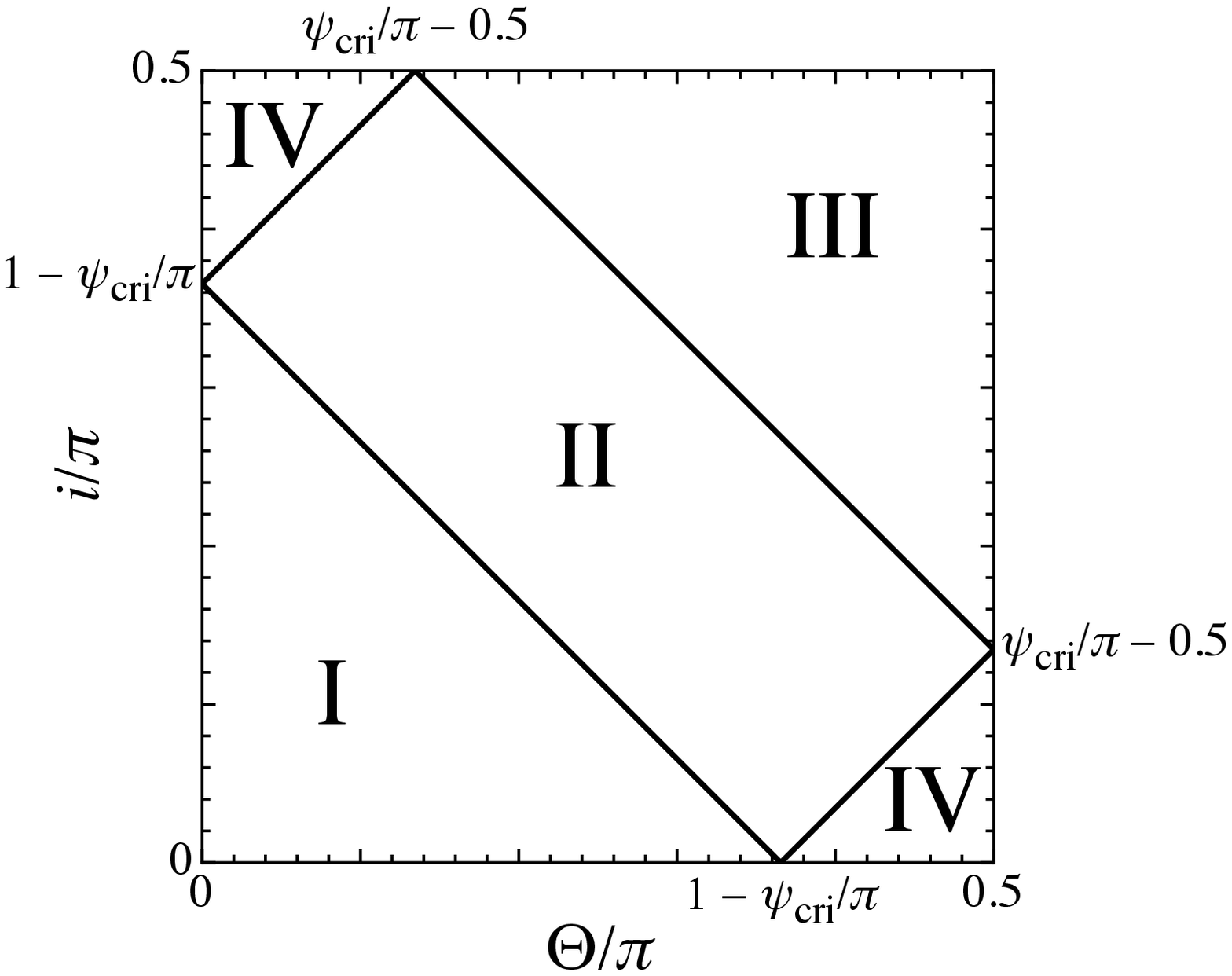} &
\includegraphics[scale=0.5]{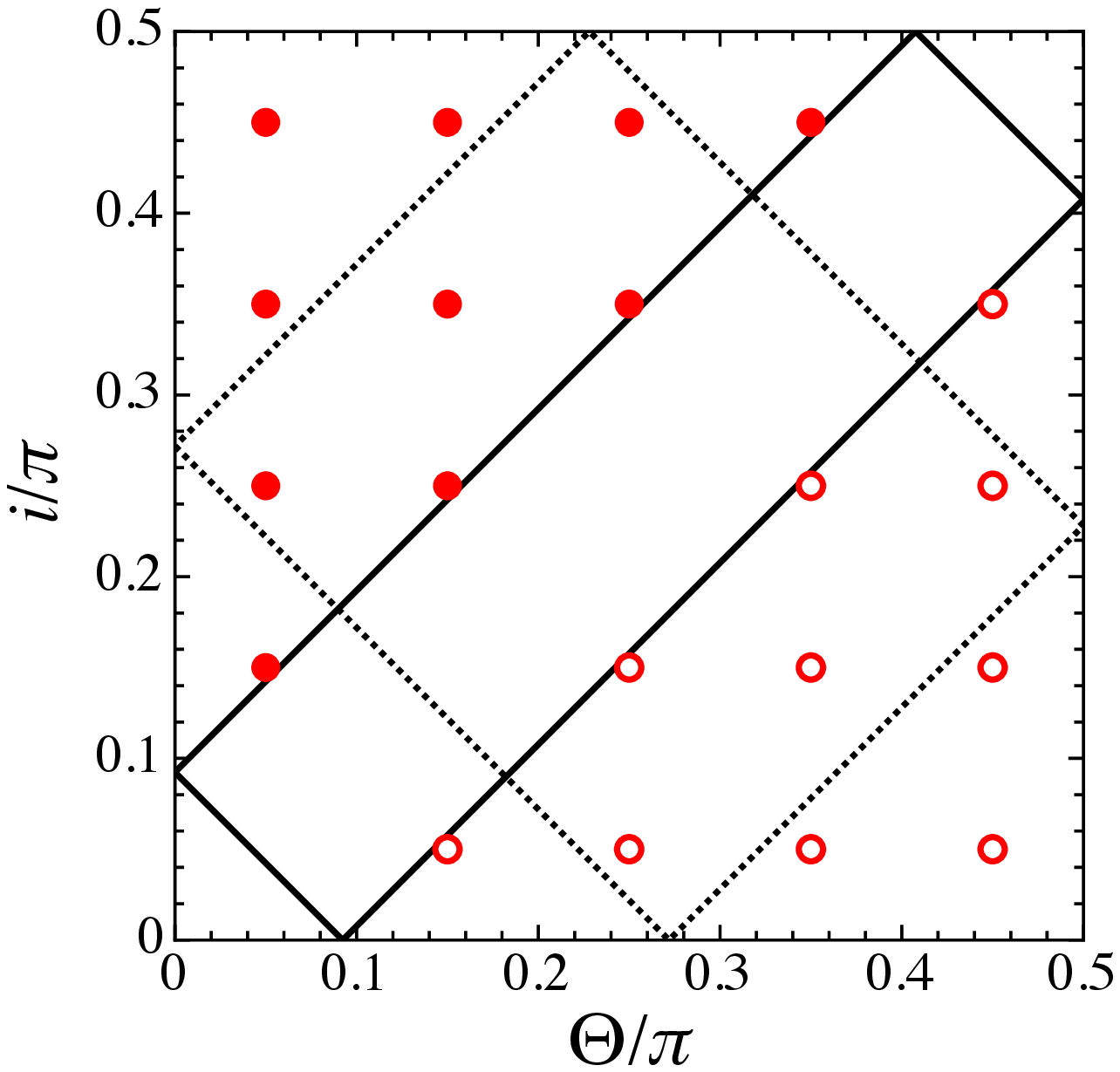} 
\end{tabular}
\end{center}
\caption{
In the left panel, the classification how the hot spots can be observed is shown with a specific value of $\psi_{\rm cri}$ \cite{SM2017}. The regions denoted with I, II, III, and IV correspond to the cases when   In the right panel, the boundaries of the classification how the hot spots can be observed are shown for the stellar models with $M/R_c=0.2658$ (solid line) and 0.2045 (dotted line). The open and filled circles denote the angles of $\Theta/\pi$ and $i/\pi$ adopted in this study.
}
\label{fig:class}
\end{figure}

The value of $\psi_{\rm cri}$ depends only on the stellar compactness after fixing the spacetime geometry, i.e., $\psi_{\rm cri}/\pi=0.908$ and 0.728 for the stellar models with $M/R_c=0.2658$ and $0.2045$ in the Schwarzschild spacetime. This value is a key parameter for determining the classification how the hot spots can be observed \cite{Beloborodov2002,SM2017}, depending on the angles of $(\Theta,i)$ as shown in the left panel of Fig. \ref{fig:class}. That is, the regions I, II, III, and IV correspond to the cases as follows;
\begin{description}
  \item[region I :] only the primary hot spot is always observed, i.e., the antipodal hot spot is always in the invisible zone,
  \item[region II :] the primary hot spot is always observed and the antipodal hot spot is also observed sometime,
  \item[region III :] the primary hot spot cannot be observed sometime and the antipodal hot spot is also observed sometime,
  \item[region IV :] the both hot spots are always observed.
\end{description}

In the right panel of Fig. \ref{fig:class}, the boundaries of such a classification are shown with solid and dotted lines for the cases of $M/R_c=0.2658$ and $0.2045$, respectively. In the same figure, the open and filled circles denote the specific angles of $i$ and $\Theta$, which we consider in this study. So, all cases considered for the stellar model with $M/R_c=0.2658$ are in the class of IV, while the cases for the stellar models with $M/R_c=0.2045$ are in either the classes of I, II, III, or IV. We remark that the observed bolometric flux $F(t)$ in the limit of $\delta\to 1$ is symmetric under the interchange of $i$ and $\Theta$ \cite{SM2017,SM2018}. One can see that the time delay $\Delta t$ has the same property. But the Doppler factor is not symmetric under the interchange of $i$ and $\Theta$, which breaks the symmetric property in the observed bolometric flux with the special relativistic effect. In the following, we will see the time delay, the Doppler effect, and the bolometric flux for different stellar models.

\subsection{Time delay}
\label{sec:IVa}

\begin{figure*}
\begin{center}
\begin{tabular}{c}
\includegraphics[scale=0.38]{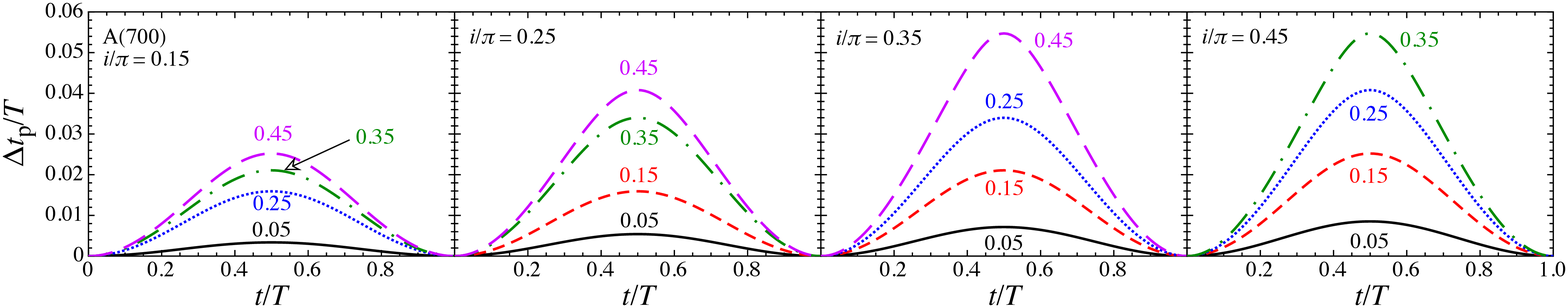} \\
\includegraphics[scale=0.38]{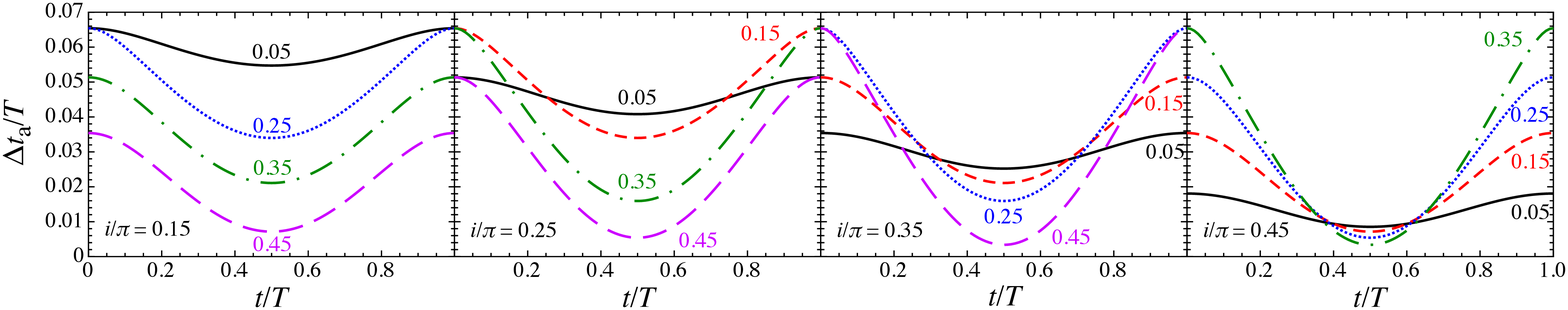} 
\end{tabular}
\end{center}
\caption{
The time delay $\Delta t_{\rm p}$ (upper panel) and $\Delta t_{\rm a}$ (lower panel) with respect to the primary and antipodal hot spots, which are normalized by the rotational period $T$, are shown as a function of $t/T$ for A(700). The panels from left to right correspond to the results with $i/\pi=0.15$, 0.25, 0.35, and 0.45. The numbers in each panel denote the value of $\Theta/\pi$.
}
\label{fig:dt-M18R10}
\end{figure*}

In Fig. \ref{fig:dt-M18R10}, the time delay $\Delta t$ normalized by the rotational period $T\equiv 2\pi/\omega$ for A(700) is shown as a function of $t/T$, where $\Delta t_{\rm p}$ and $\Delta t_{\rm a}$ denote the time delay calculated for the primal and antipodal hot spots. From this figure, one can confirm that the time delay is symmetric under the interchange of $i$ and $\Theta$. Additionally, we find that the time delay can be $\sim 7\%$ of the rotational period, which may not be negligible. However, this effect would become small as the rotational frequency decreases, because the value of the time delay is determined with the stellar radius and mass, while it is independent of the rotational frequency. In fact, the time delay becomes at most $\sim 0.1\%$ of the rotational period, considering $\nu=10$ Hz, where the time delay is almost negligible for determining the pulse profile. We also remark that $\Delta t_p$ becomes zero at $t/T=0$ (and 1) by definition. On the other hand, $\Delta t_a$ does not become zero but maximum at $t/T=0$ (and 1), where the position of the antipodal hot spot becomes farthermost from the observer, and never becomes zero in the rotational period.

\begin{figure*}
\begin{center}
\begin{tabular}{cc}
\includegraphics[scale=0.5]{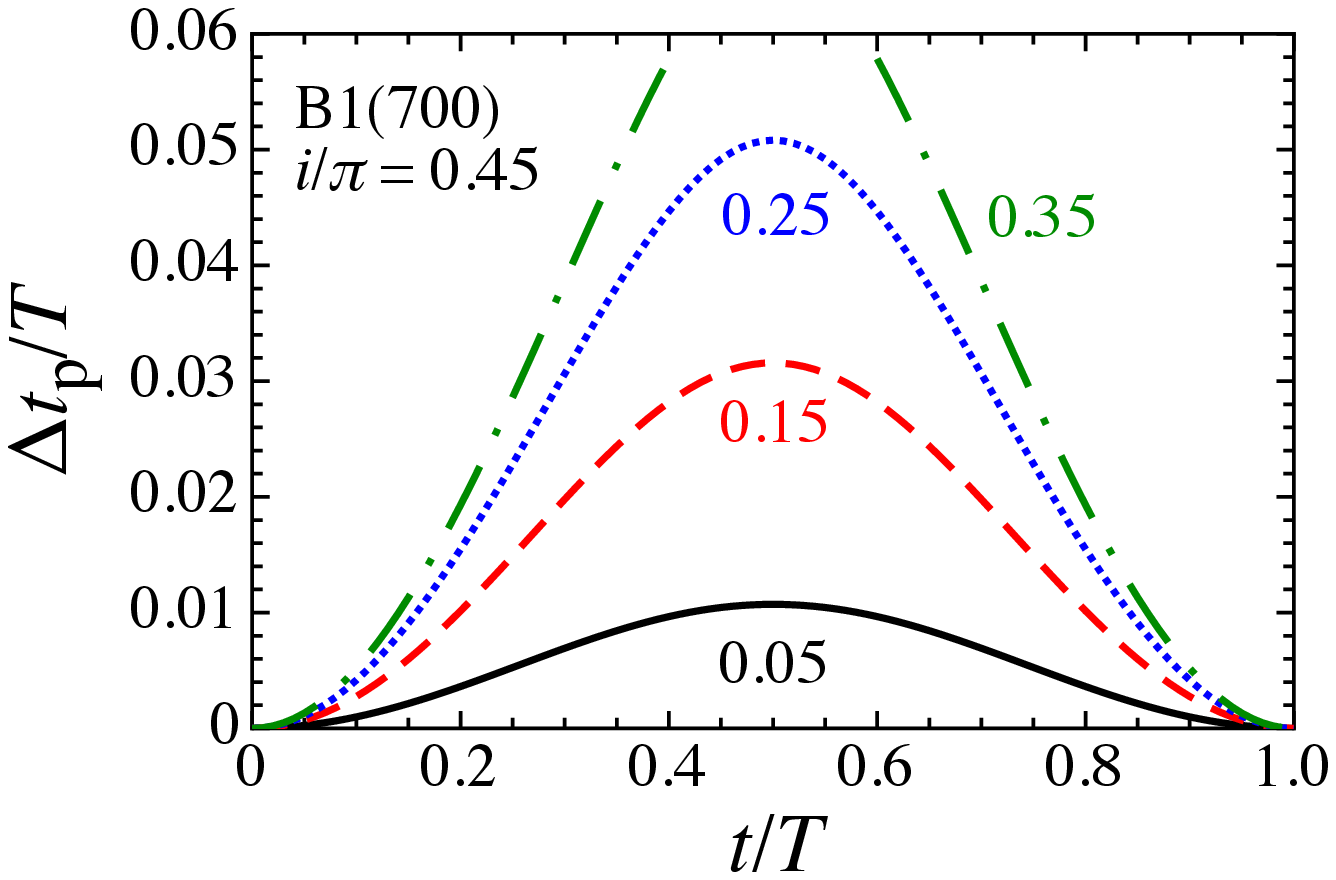} &
\includegraphics[scale=0.5]{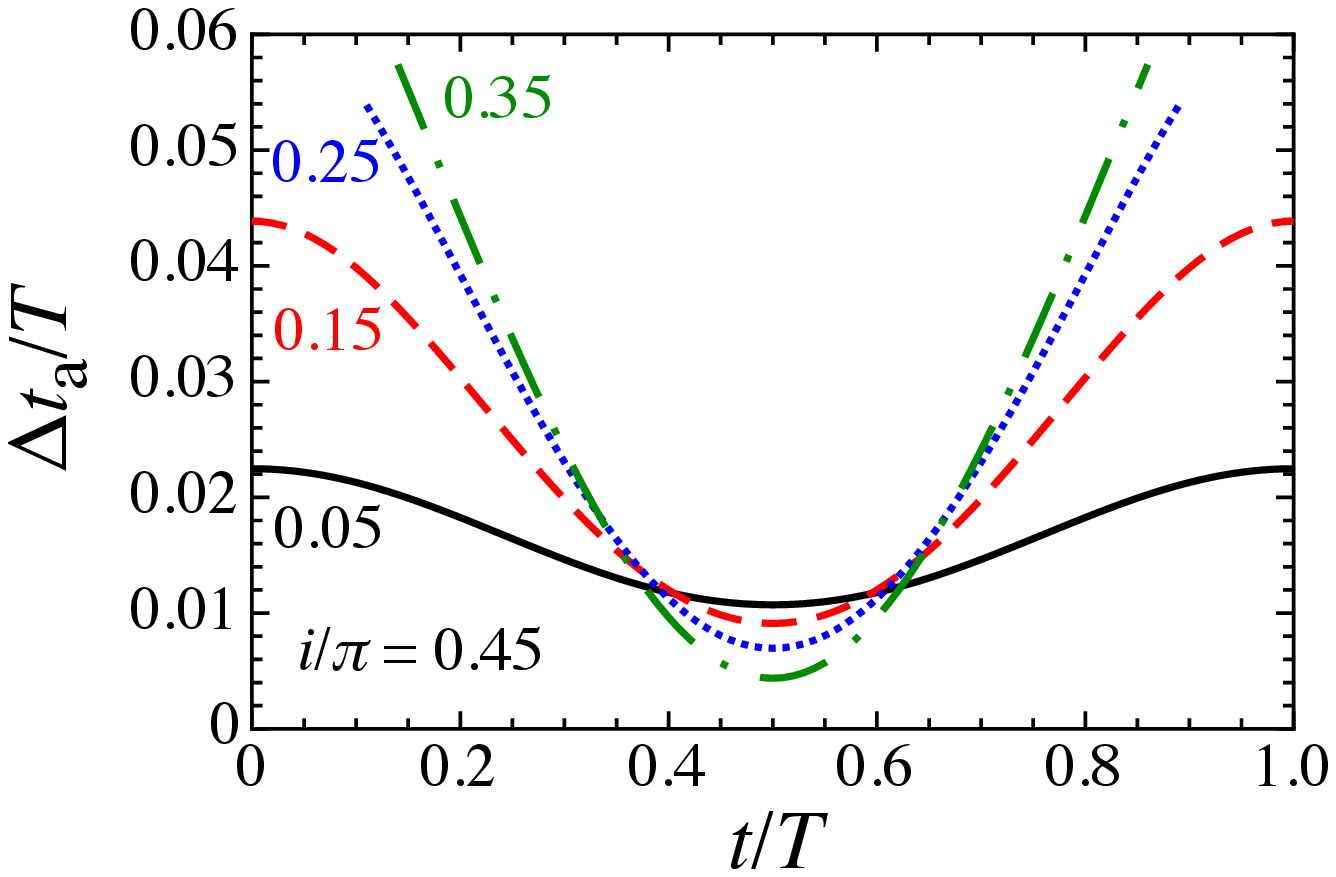} 
\end{tabular}
\end{center}
\caption{
$\Delta t_{\rm p}/T$ and $\Delta t_{\rm a}/T$ for B1(700) with $i/\pi=0.45$, where the numbers denote the value of $\Theta/\pi$.
}
\label{fig:dt-M18R13}
\end{figure*}

\begin{figure*}
\begin{center}
\begin{tabular}{cc}
\includegraphics[scale=0.5]{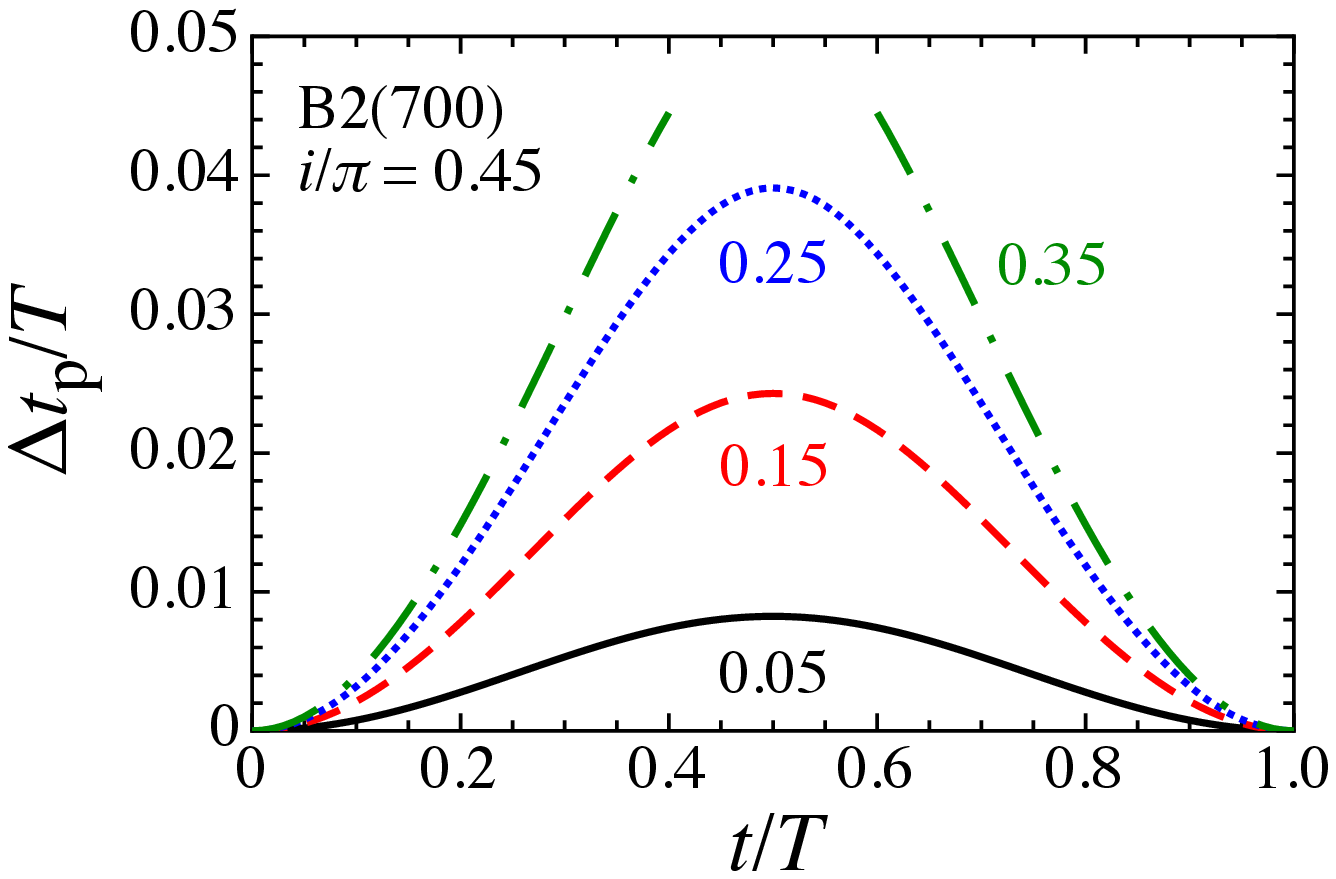} &
\includegraphics[scale=0.5]{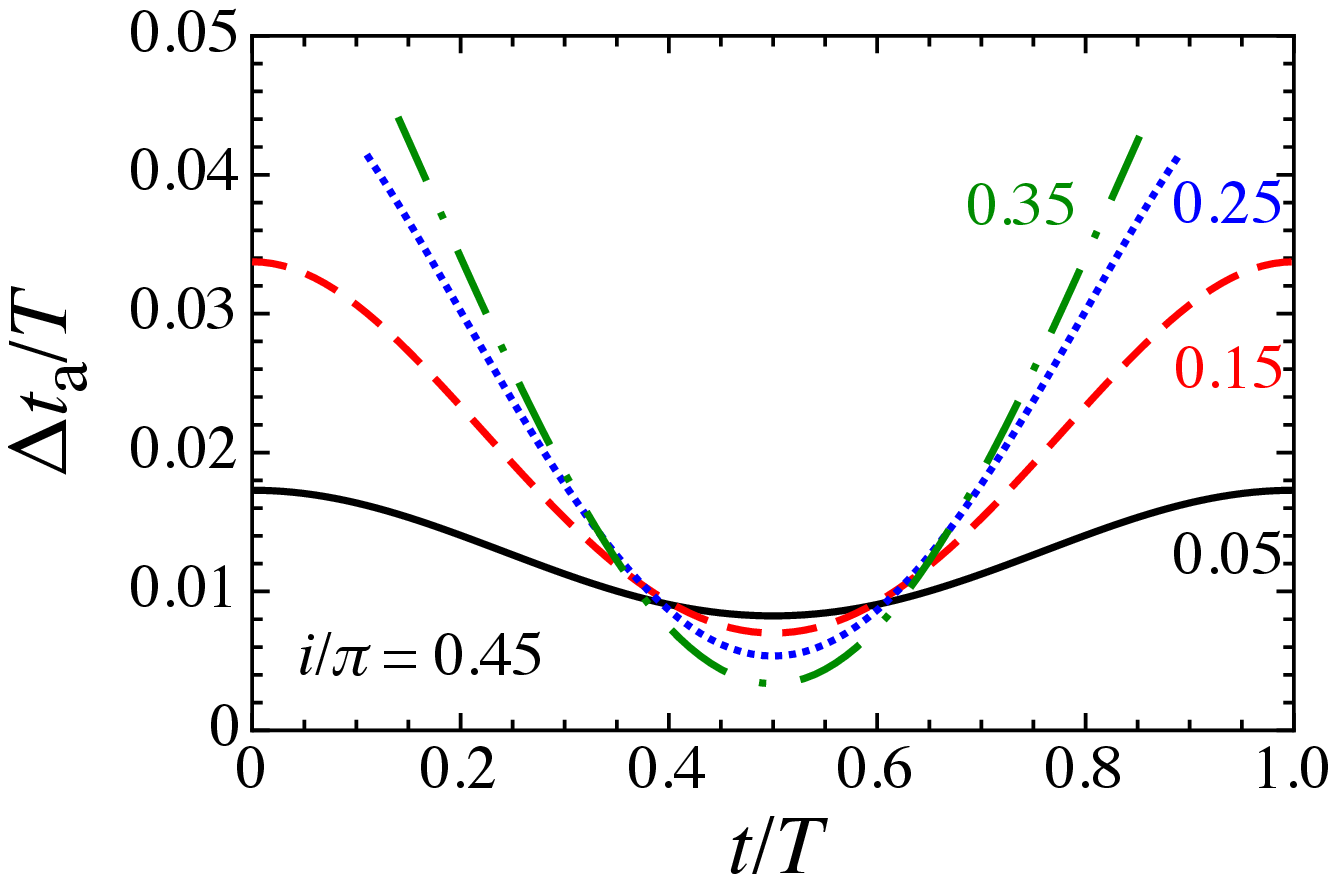} 
\end{tabular}
\end{center}
\caption{
Same as Fig. \ref{fig:dt-M18R13} but for B2(700).
}
\label{fig:dt-MR02045}
\end{figure*}

We show the time delay $\Delta t_{\rm p}$ and $\Delta t_{\rm a}$ for B1(700) in Fig. \ref{fig:dt-M18R13} and for B2(700) in Fig. \ref{fig:dt-MR02045}, where the inclination angle is fixed to be $i/\pi=0.45$. In these cases, the hot spots may enter the invisible zone, depending on the combination of the angles of $i$ and $\Theta$. In these figures, the endpoints of some lines correspond to the incursion into or the escape from the invisible zone. In addition, it is noticed that the observed bolometric flux $F(t)$ in the limit of $\delta\to 1$ depends only on the stellar compactness \cite{SM2017,SM2018}, i.e., $F(t)$ in the limit of $\delta \to 1$ for the model B1 should be the same as for the model B2. On the other hand, as shown in Figs. \ref{fig:dt-M18R13} and \ref{fig:dt-MR02045}, the time delay is not determined only with the stellar compactness, where it also depends on the stellar radius. That is, even if the stellar compactness is the same, the time delay becomes longer as the stellar radius is larger. Thus, if the neutron star rotates so fast that the effect of the time delay cannot be negligible, one might obtain the information about radius via carefully observing the pulse profile. Anyway, since the compactness of the stellar models shown in Figs. \ref{fig:dt-M18R13} and \ref{fig:dt-MR02045} is lower than that shown in Fig. \ref{fig:dt-M18R10}, the time delay in Figs. \ref{fig:dt-M18R13} and \ref{fig:dt-MR02045} becomes shorter than that in Fig. \ref{fig:dt-M18R10}. Anyway, as for the model A(700) shown in Fig. \ref{fig:dt-M18R10}, $\Delta t_a$ is always positive and becomes maximum at $t/T=0$ (and 1) or when the antipodal hot spot exits from and enters in the invisible zone.

\subsection{Doppler factor}
\label{sec:IVb}

\begin{figure*}
\begin{center}
\begin{tabular}{c}
\includegraphics[scale=0.38]{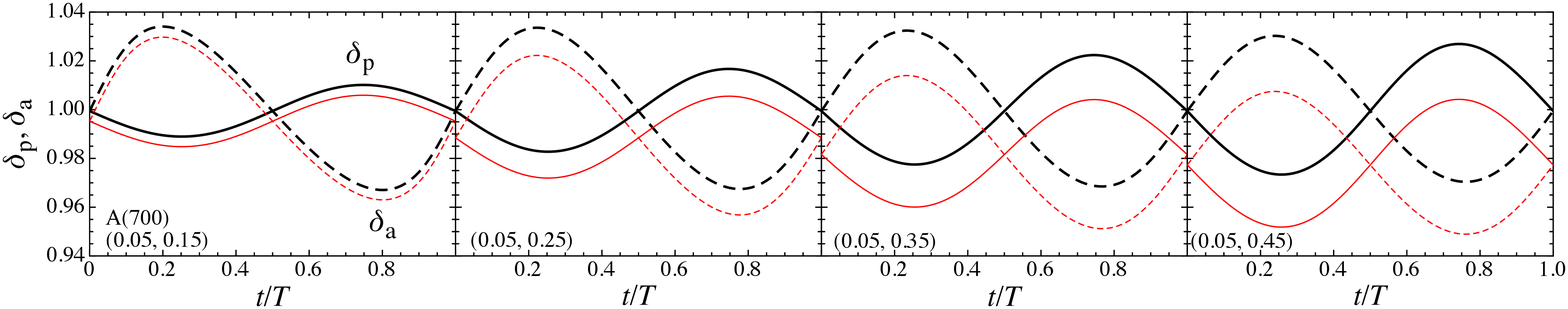} \\
\includegraphics[scale=0.38]{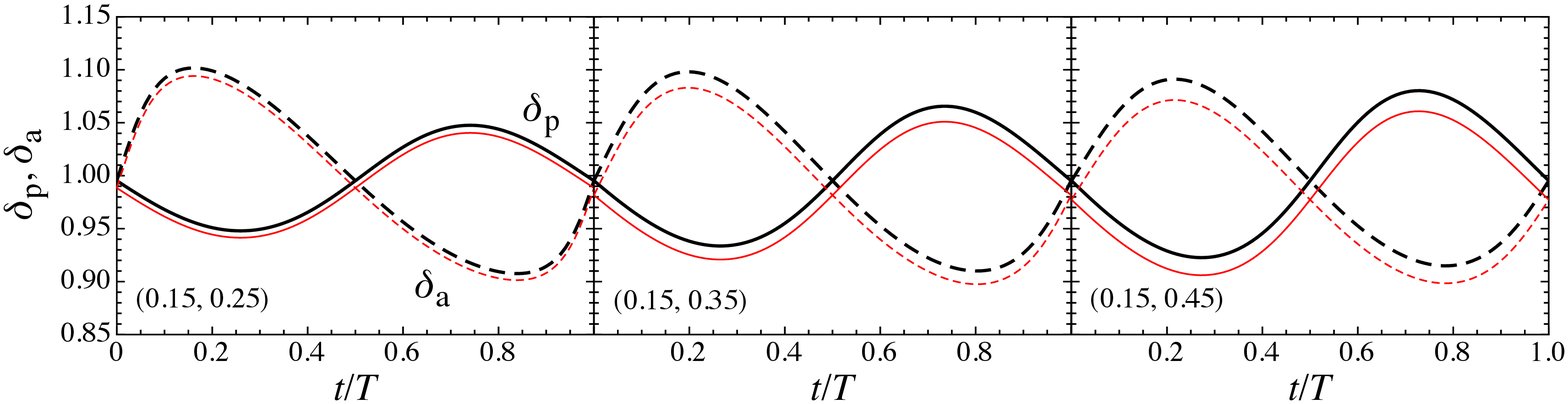} \\ 
\includegraphics[scale=0.38]{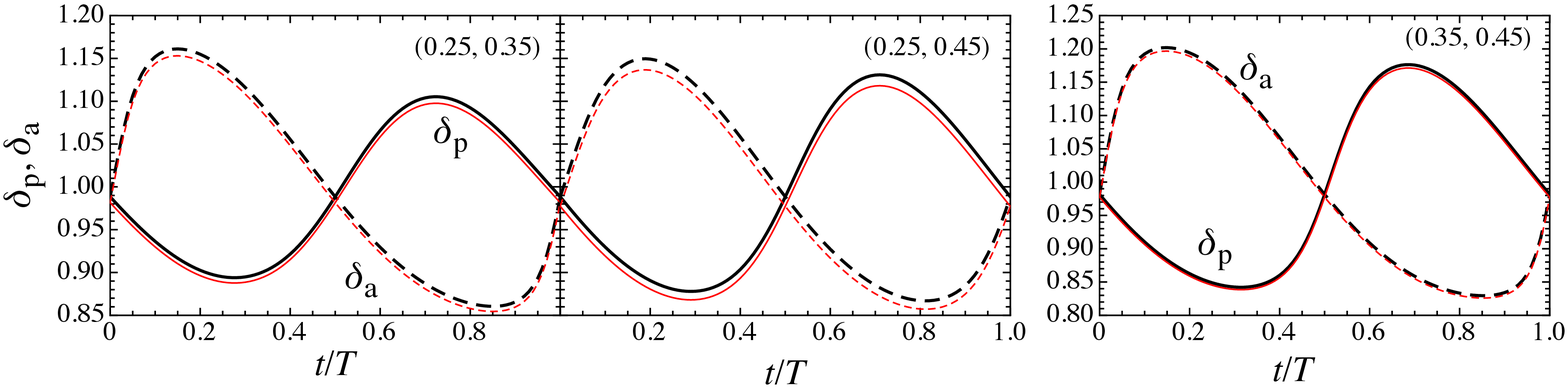} 
\end{tabular}
\end{center}
\caption{
Doppler factor $\delta$ is shown as a function of $t/T$, where $\delta_{\rm p}$ (solid line) and $\delta_{\rm a}$ (dashed line) denote the value of $\delta$ for the primary and antipodal hot spots, for A(700). In each panel, $(a,b)$ denotes $(i/\pi$, $\Theta/\pi)$ or $(\Theta/\pi, i/\pi)$: the thin lines correspond to the results with $(i/\pi$, $\Theta/\pi)=(a,b)$; the thick lines correspond to those with $(i/\pi$, $\Theta/\pi)=(b,a)$. 
}
\label{fig:delta-M18R10}
\end{figure*}

\begin{figure}
\begin{center}
\includegraphics[scale=0.5]{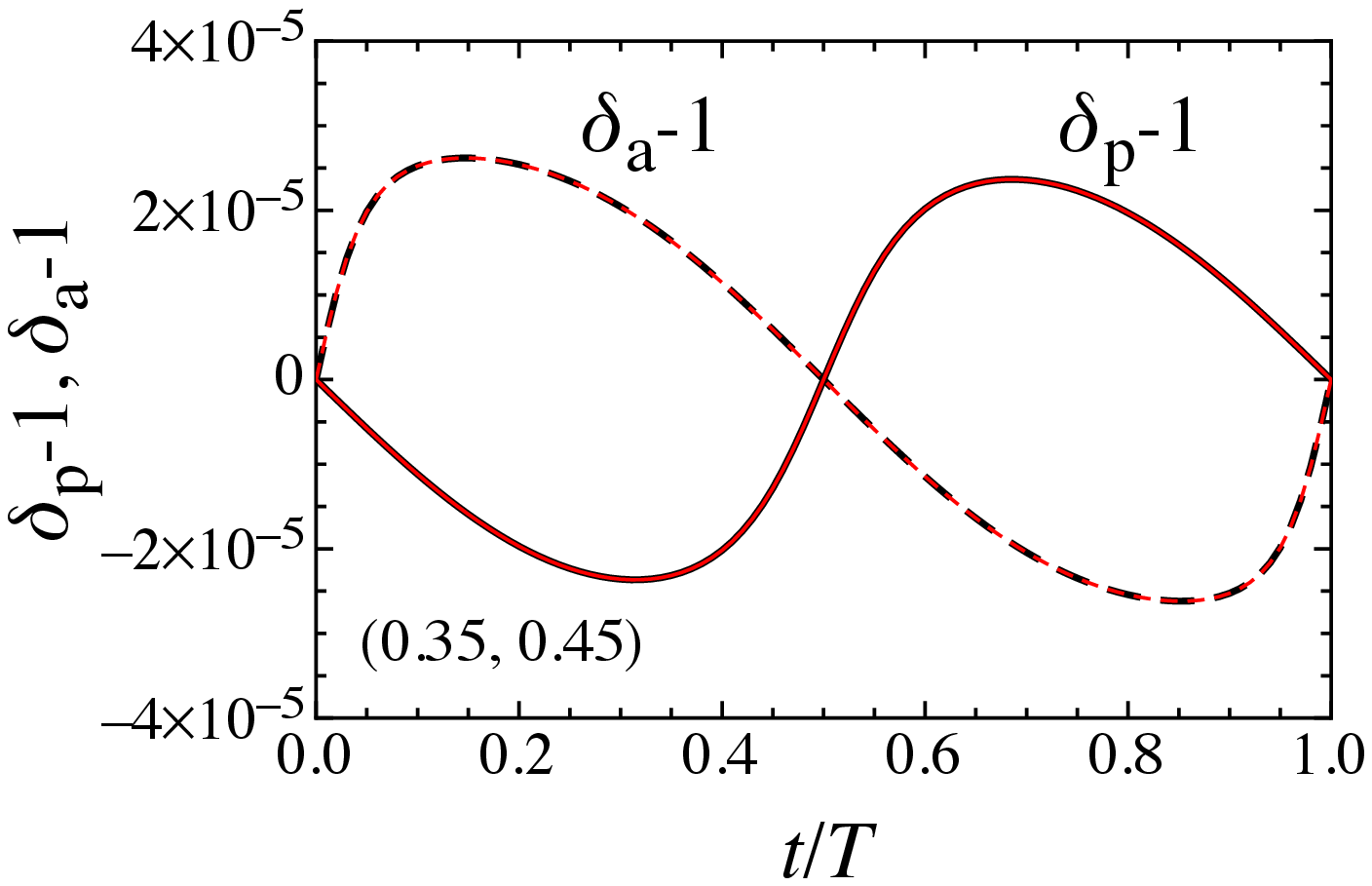} 
\end{center}
\caption{
For A(0.1), $\delta_{\rm p}-1$ and $\delta_{\rm a}-1$ are shown as a function of $t/T$ with $(i/\pi,\Theta/\pi)=(0.35, 0.45)$ and $(0.45, 0.35)$.
}
\label{ig:delta-M18R10-f01}
\end{figure}

In Fig. \ref{fig:delta-M18R10}, the Doppler factor $\delta$ for A(700) with various angles of $i$ and $\Theta$ is shown as a function of $t/T$, where $\delta_{\rm p}$ (solid lines) and $\delta_{\rm a}$ (dashed lines) denote the Doppler factor for the primary and antipodal hot spots. In each panel, $(a,b)$ denotes the angles $(i/\pi,\Theta/\pi)$ or $(\Theta/\pi, i/\pi)$:  the thin lines correspond to the results with $(i/\pi,\Theta/\pi)=(a,b)$; the thick lines correspond to results with $(i/\pi,\Theta/\pi)=(b,a)$. Unlike the case of $F(t)$ in the limit of $\delta \to 1$ and $\Delta t$, the values of $\delta$ are not symmetric under the interchange of $i$ and $\Theta$. In fact, one can see that the difference between $\delta_{\rm p}$ and $\delta_{\rm a}$ increases as the value of $|i-\Theta|$ increases. For any combination of $i$ and $\Theta$, we also find that the value of $\delta$ with $i<\Theta$ is smaller than that with $i>\Theta$. In Fig. \ref{ig:delta-M18R10-f01}, we show $\delta_{\rm p}-1$ and $\delta_{\rm a}-1$ for A(0.1). From this figure, one can see that the effect of the Doppler factor is negligible for the stellar model with slow rotation such as a magnetar.

\begin{figure*}
\begin{center}
\begin{tabular}{cc}
\includegraphics[scale=0.5]{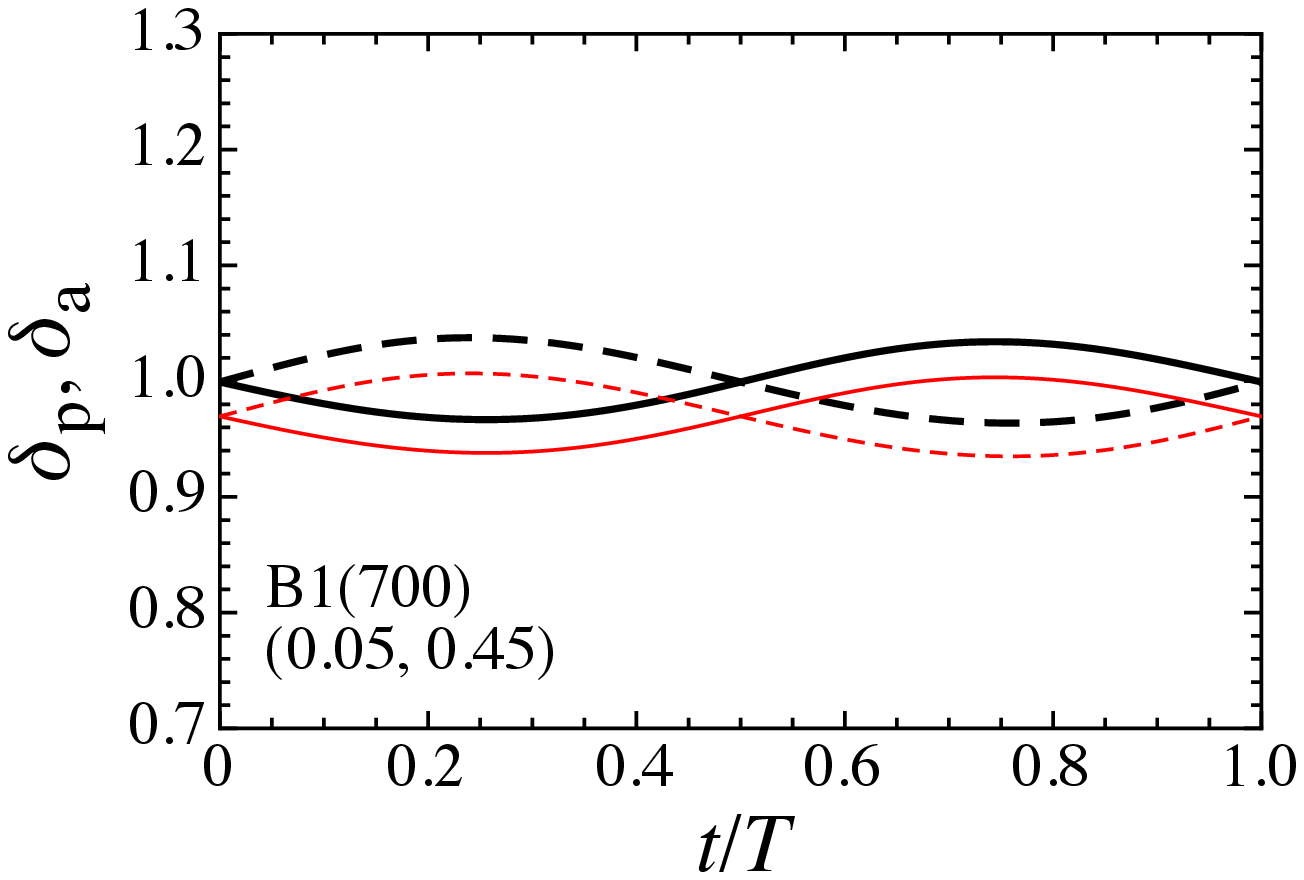} &
\includegraphics[scale=0.5]{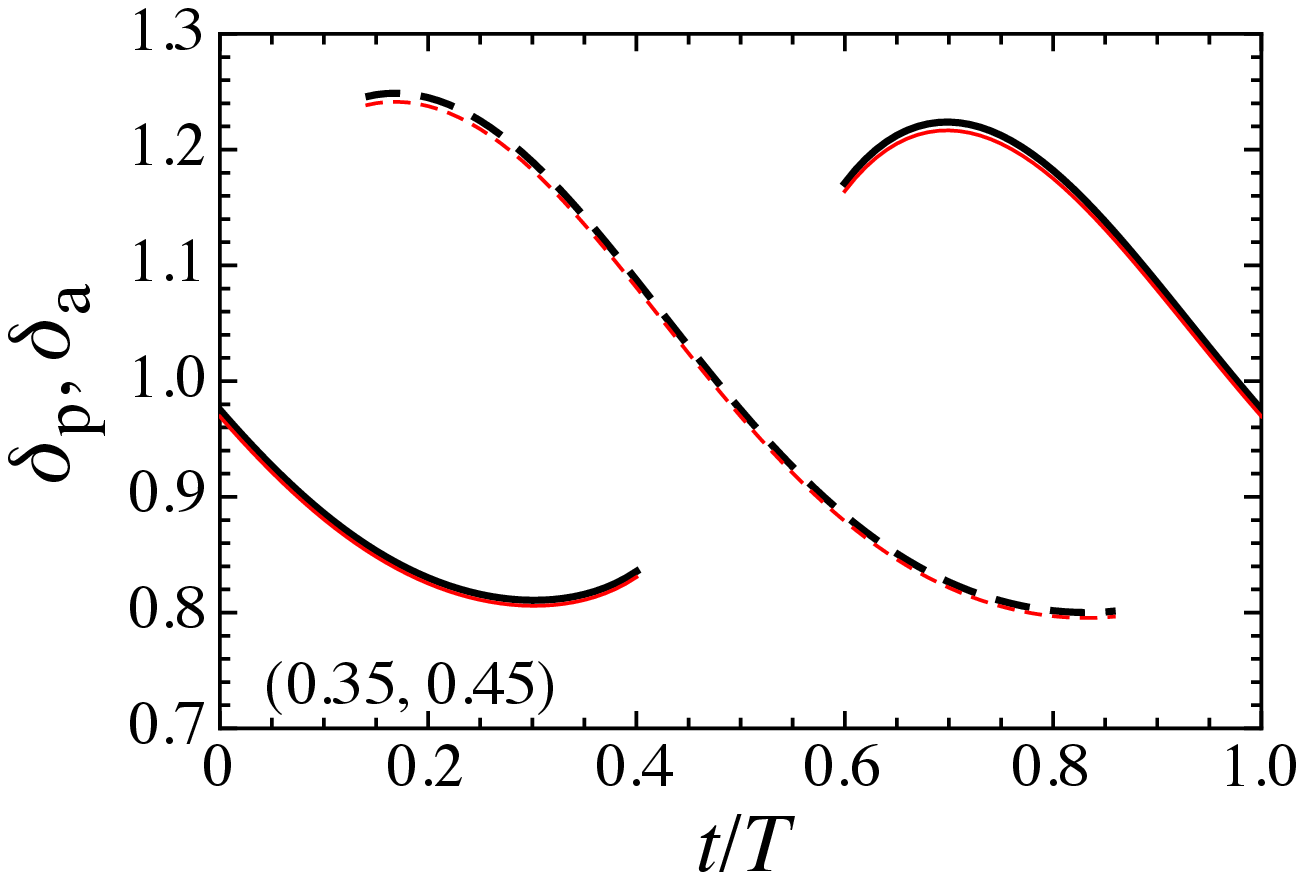} 
\end{tabular}
\end{center}
\caption{
For B1(700), $\delta_{\rm p}$ and $\delta_{\rm a}$ are shown as a function of $t/T$, adopting the combinations of angles of $(i/\pi,\Theta/\pi)=(0.05, 0.45)$ and $(0.45, 0.05)$ in the left panel and those of $(0.35, 0.45)$ and $(0.45, 0.35)$ in the right panel. In the right panel, we show only when the hot spots can be observed. 
}
\label{fig:delta-M18R13}
\end{figure*}

\begin{figure*}
\begin{center}
\begin{tabular}{cc}
\includegraphics[scale=0.5]{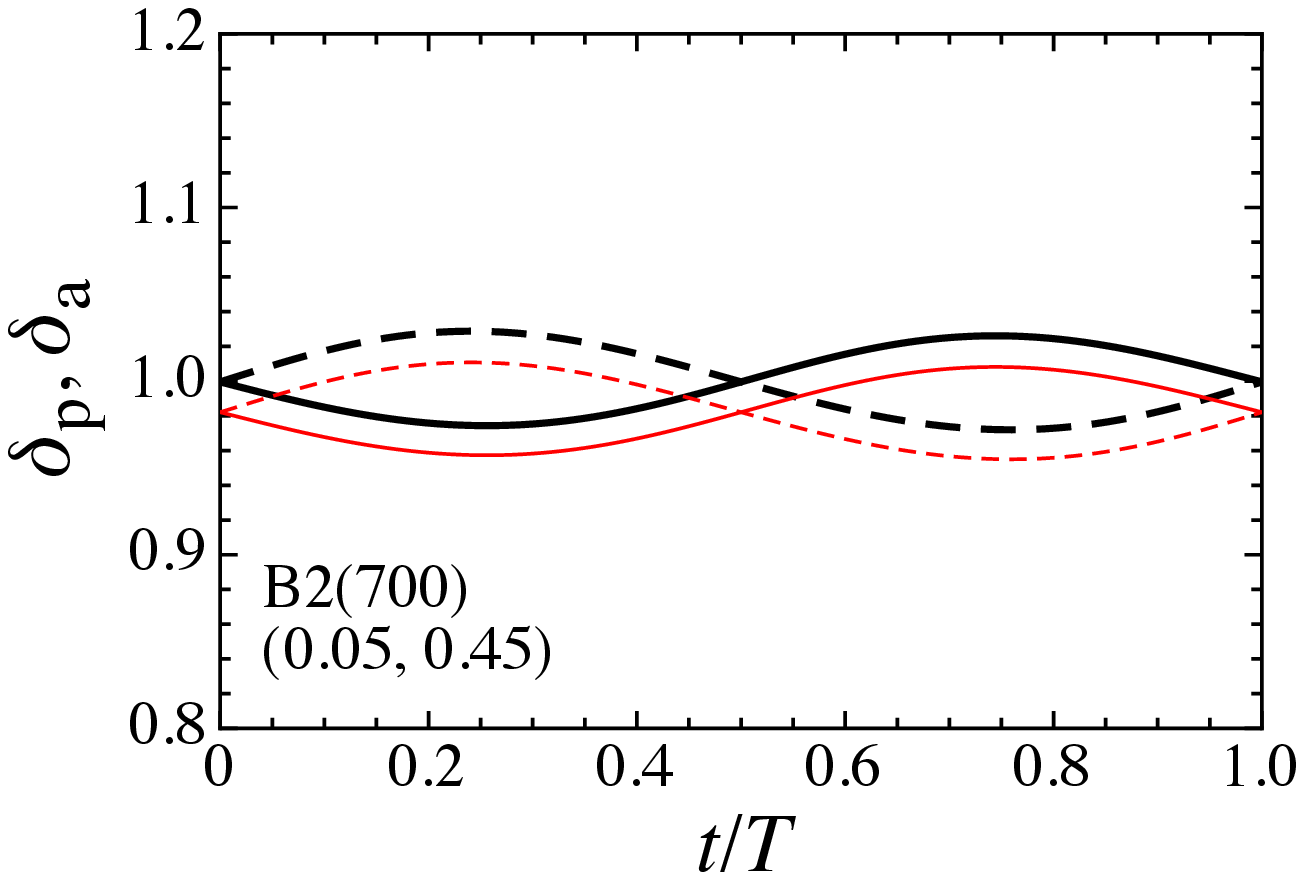} &
\includegraphics[scale=0.5]{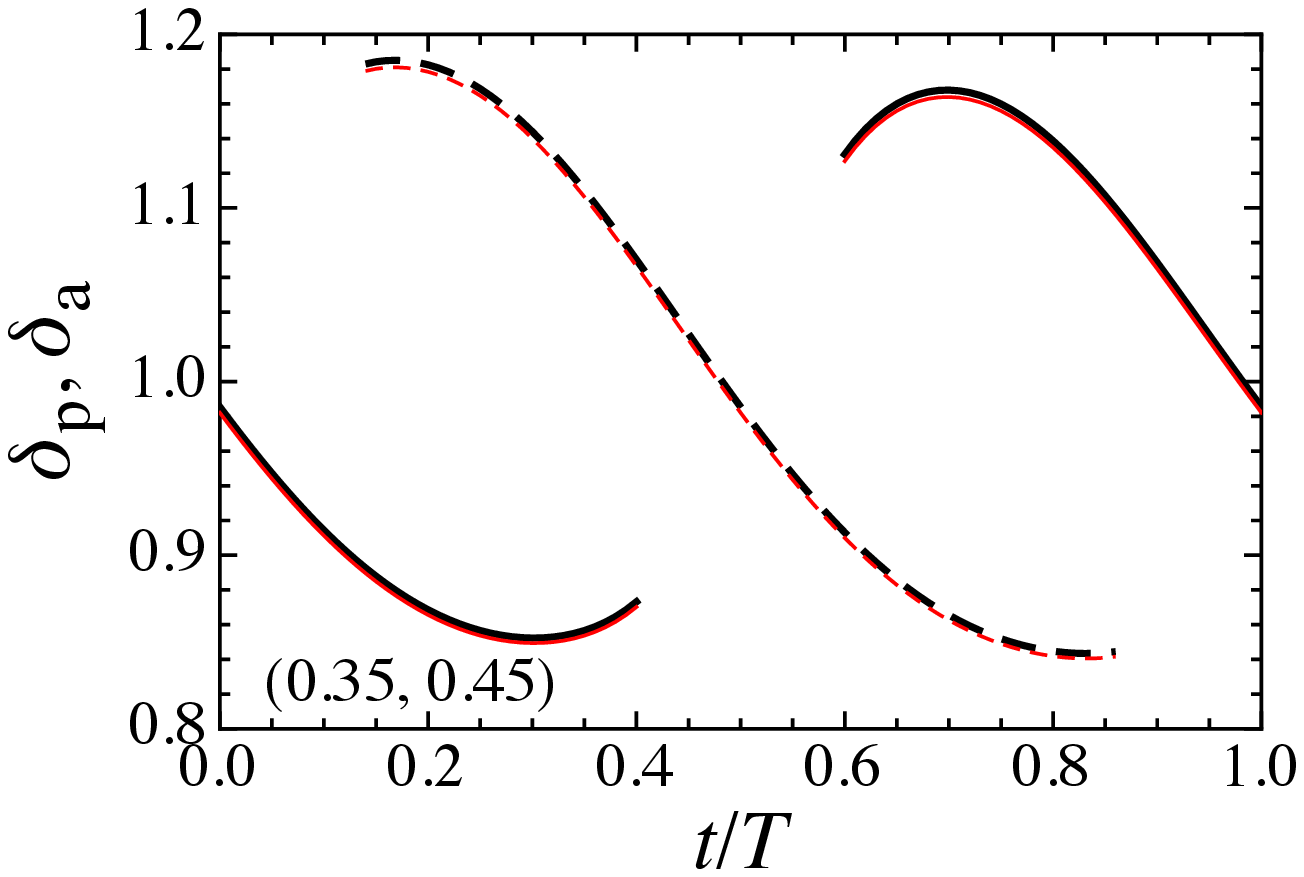} 
\end{tabular}
\end{center}
\caption{
Same as Fig. \ref{fig:delta-M18R13} but for B2(700).
}
\label{fig:delta-MR02045}
\end{figure*}

We also show the values of $\delta_{\rm p}$ and $\delta_{\rm a}$ as a function of $t/T$ for B1(700) in Fig. \ref{fig:delta-M18R13} and for B2(700) in Fig. \ref{fig:delta-MR02045}. In both figures, we particularly adopt the combinations of angles of $(i/\pi,\Theta/\pi)=(0.05, 0.45)$ and $(0.45, 0.05)$ in the left panel and those of $(0.35, 0.45)$ and $(0.45, 0.35)$ in the right panel. From these figures, we can observe that the shape of the Doppler factor as a function of $t/T$ shown in Fig. \ref{fig:delta-M18R13} is the same as that shown in Fig. \ref{fig:delta-MR02045}, but the absolute value of it depends on the stellar radius, i.e., the absolute value of the Doppler factor becomes larger as the stellar radius is larger even if the stellar compactness is the same. That is, the pulse profiles depend not only the stellar compactness but also the stellar radius, if the neutron star is fast rotating.

\subsection{Observed bolometric flux}
\label{sec:IVc}

\begin{figure*}
\begin{center}
\begin{tabular}{c}
\includegraphics[scale=0.38]{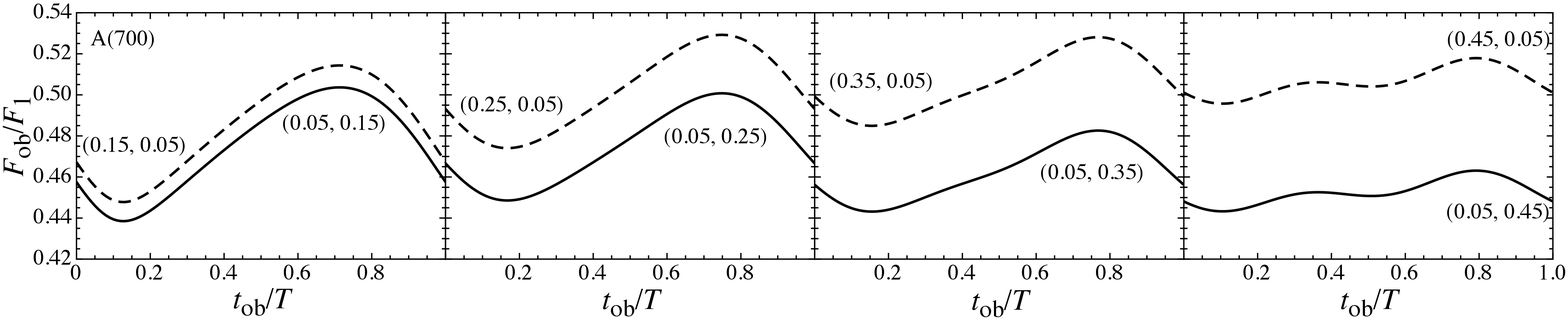} \\
\includegraphics[scale=0.38]{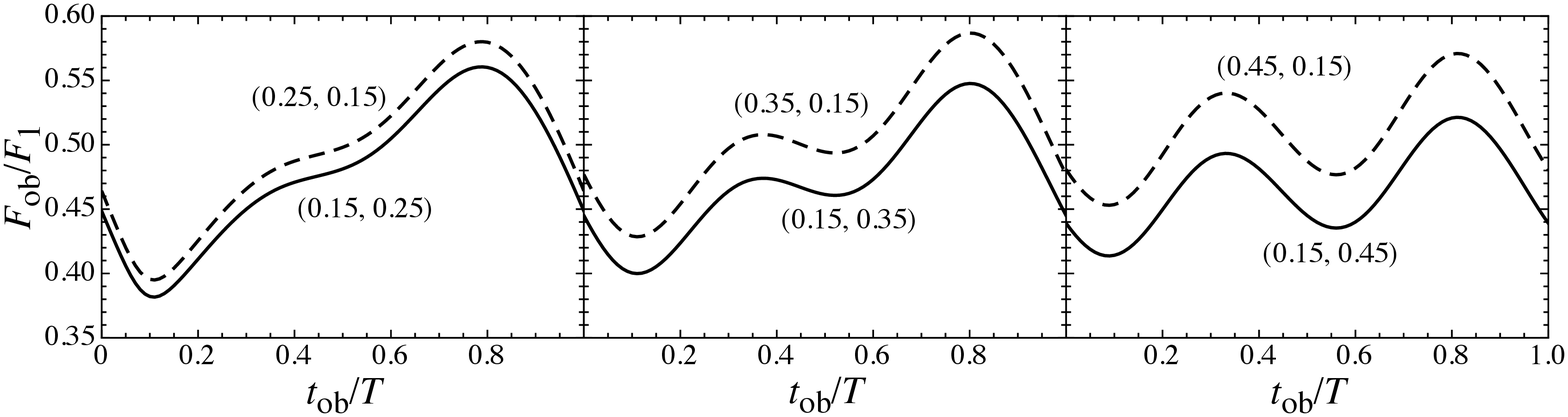} \\ 
\includegraphics[scale=0.38]{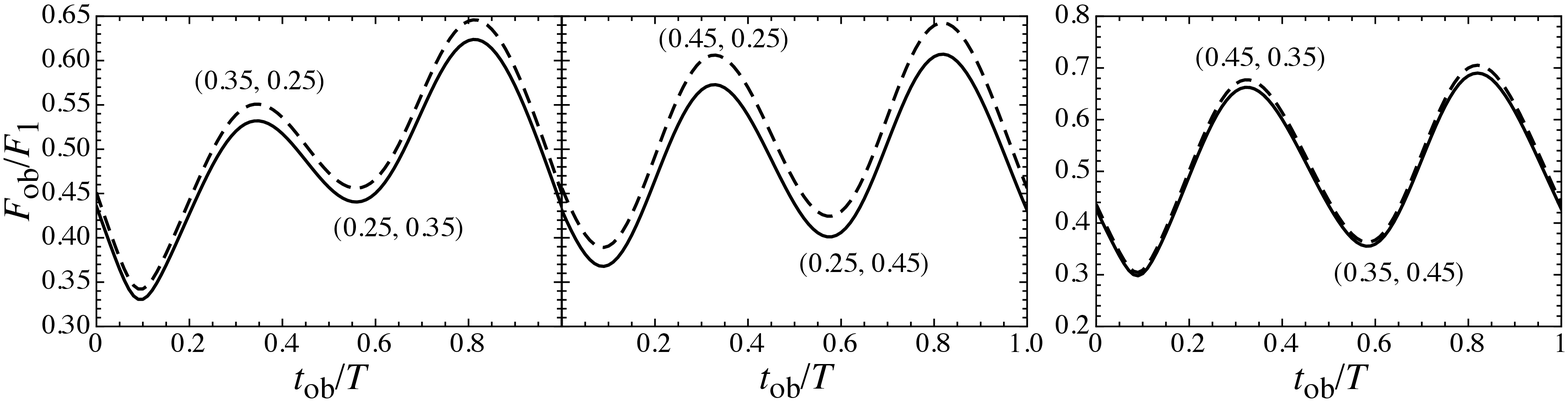} 
\end{tabular}
\end{center}
\caption{
The observed bolometric flux $F_{\rm ob}$ normalized by $F_1$ are shown as a function of $t/T$ for A(700), where in each panel $(*,*)$ denotes the value of $(i/\pi,\Theta/\pi)$. The solid and dashed lines denote the results with $(i/\pi,\Theta/\pi)=(a,b)$ and $(b,a)$ for $a<b$, respectively.
}
\label{fig:Fob-M18R10nu700}
\end{figure*}

\begin{figure}
\begin{center}
\includegraphics[scale=0.5]{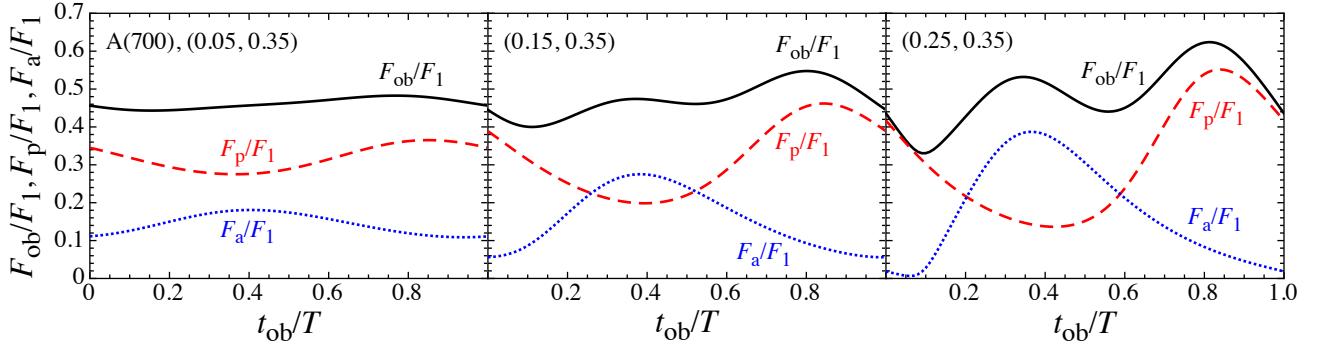} 
\end{center}
\caption{
For A(700), the observed bolometric flux $F_{\rm ob}/F_1$, the flux from the primary spot $F_{\rm p}/F_1$, and the flux from the antipodal sopt $F_{\rm a}/F_1$ are shown with the solid, dashed, and dotted lines for $(i/\pi,\Theta/\pi)=(0.05, 0.35)$ in the left panel, (0.15, 0.35) in the middle panel, and (0.25,0.35) in the right panel.
}
\label{fig:dtob-A700FobFpFa}
\end{figure}

\begin{figure}
\begin{center}
\includegraphics[scale=0.5]{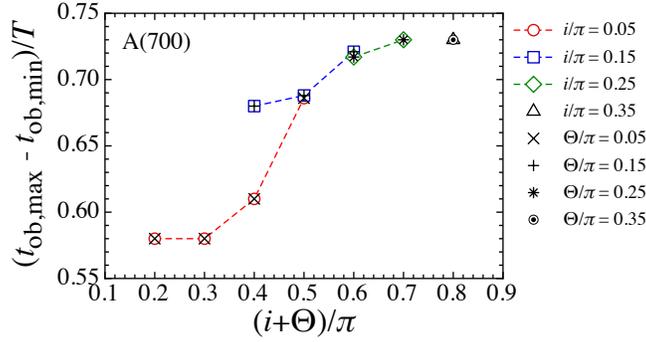} 
\end{center}
\caption{
For A(700), the interval between the minimum and the maximum in the light curve are shown as a function of $(i+\Theta)/\pi$ for various combinations of the angles of $i$ and $\Theta$.
}
\label{fig:dtob-M18R10nu700}
\end{figure}

The observed bolometric flux from the primary and antipodal hot spots can be calculated with Eq. (\ref{eq:FF1}). Taking into account the time delay, one can transform it to those as a function of $t_{\rm ob}$. By adding the observed bolometric flux from the both hot spots, the observed bolometric flux, $F_{\rm ob}$, is finally obtained. First, we consider the pulse profiles for A(700). With various angles of $i$ and $\Theta$ shown in the right panel of Fig. \ref{fig:class}, the profiles of $F_{\rm ob}/F_1$ are shown as a function of $t_{\rm ob}/T$ in Fig. \ref{fig:Fob-M18R10nu700}, where the angles of ($i/\pi$, $\Theta/\pi$) are shown in each panel, and the solid and dashed lines denote the results with $(i/\pi,\Theta/\pi)=(a,b)$ and $(b,a)$ for $a<b$, respectively. From this figure, one can observe that the shape of pulse profile with $(i/\pi,\Theta/\pi)=(a,b)$ is the same as that with  $(i/\pi,\Theta/\pi)=(b,a)$, where the just amplitude is different from each other (see Appendix. \ref{sec:a1}). This is because the values of $F_{\rm ob}/F_1(t_{\rm ob})$ in the limit of $\delta\to 1$ and $t_{\rm ob}$ are symmetric under the interchange of $i$ and $\Theta$ but $\delta$ is not symmetric. Thus, the difference of the amplitude in each panel of Fig. \ref{fig:Fob-M18R10nu700} comes from the asymmetry of $\delta$ under the interchange of $i$ and $\Theta$. In addition, from Fig. \ref{fig:Fob-M18R10nu700} one can observe the appearance of the local maxima and local minima between the minimum and maximum of the flux in the light curve, if the value of $(i+\Theta)/\pi$ is larger than $\sim 0.5$, and that the absolute values of local maxima and local minima increase as $(i+\Theta)/\pi$ increases. 
To clearly see this behavior, we show, in Fig. \ref{fig:dtob-A700FobFpFa}, the observed bolometric flux $F_{\rm ob}/F_1$, the flux from the primary spot $F_{\rm p}/F_1$, and the flux from the antipodal sopt $F_{\rm a}/F_1$ with the solid, dashed, and dotted lines, where the left, middle, and right panels correspond to $(i/\pi,\Theta/\pi)=(0.05, 0.35)$, (0.15, 0.35), and (0.25,0.35). From this figure, one can observe that the appearance of the local maxima and local minima, and their amplitudes strongly depend on the amplitudes of $F_{\rm p}/F_1$ and $F_{\rm a}/F_1$. In fact, as $(i+\Theta)/\pi$ increases, the change of the distance between the both hot spots and the observer can be larger, which leads to the results that the amplitudes of $F_{\rm p}/F_1$ and $F_{\rm a}/F_1$ become larger. As a result, the local maxima and local minima appear and their amplitudes become larger, as $(i+\Theta)/\pi$ increases.
One can also observe that the interval between the minimum and maximum in light curve increases as $(i+\Theta)/\pi$ increases, as shown in Fig. \ref{fig:dtob-M18R10nu700}, where $t_{\rm ob,min}/T$ and $t_{\rm ob,max}/T$ denote the values of $t_{\rm ob}/T$ in Fig. \ref{fig:Fob-M18R10nu700} when the observed flux becomes the minimum and the maximum, respectively. Anyway, taking into account that such an interval should be 0.5 in the slow rotation limit, one can say that the modification of light curve in any case due to the fast rotation occurs. 

\begin{figure}
\begin{center}
\includegraphics[scale=0.5]{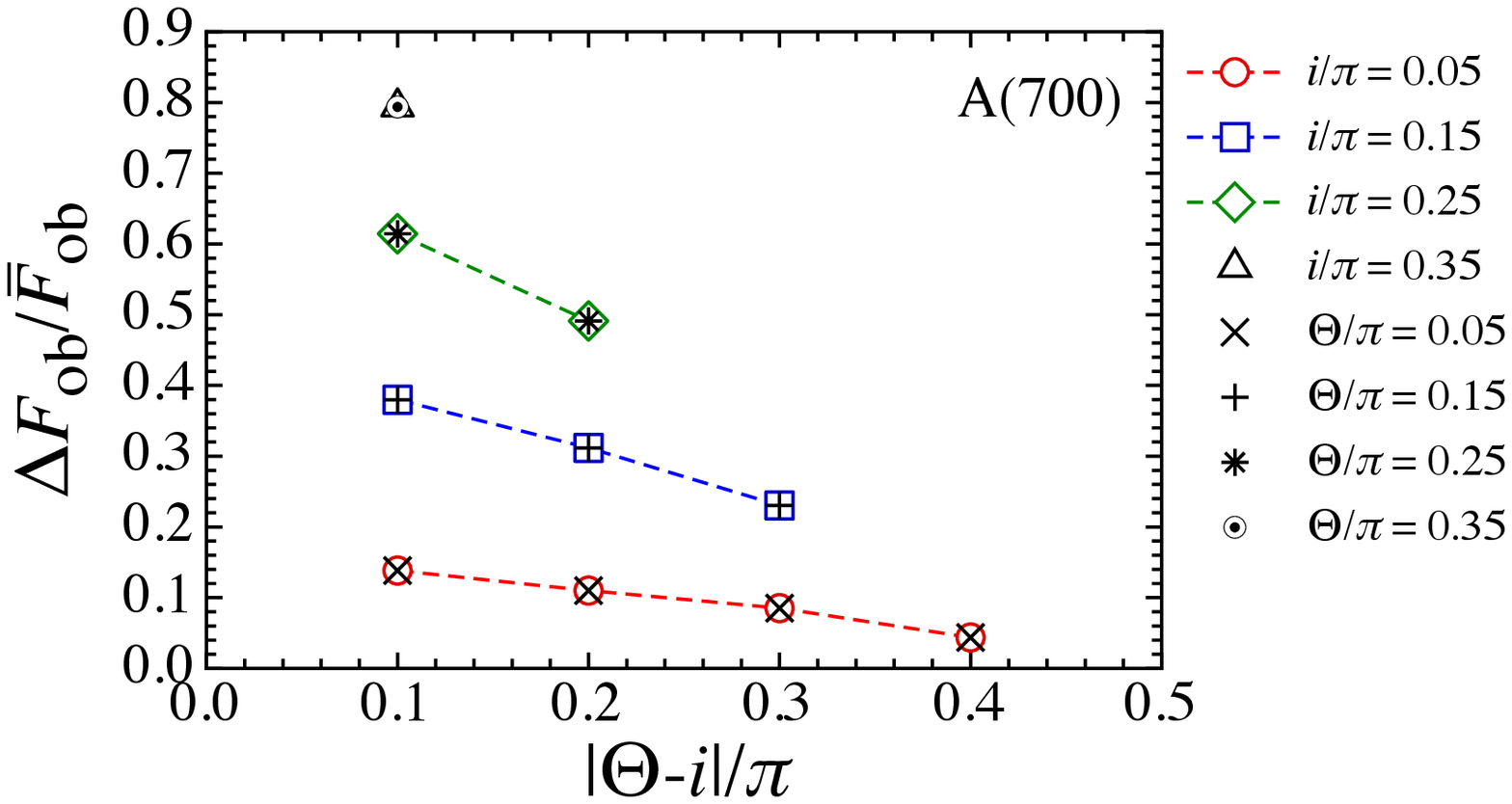} 
\end{center}
\caption{
$\Delta F_{\rm ob}/\bar{F}_{\rm ob}$ for A(700) is shown as a function of $|\Theta-i|/\pi$, where $\Delta F_{\rm ob}\equiv F_{\rm ob}^{(\rm max)}-F_{\rm ob}^{(\rm min)}$ and $\bar{F}_{\rm ob}\equiv (F_{\rm ob}^{(\rm max)}+F_{\rm ob}^{(\rm min)})/2$ with the maximum and minimum values of $F_{\rm ob}$, $F_{\rm ob}^{(\rm max)}$ and $F_{\rm ob}^{(\rm min)}$. In the figure, the open circles, squares, diamonds, and triangle denote the results with $i/\pi=0.05$, 0.15, 0.25, and 0.35, while the crosses, pluses, asterisks, and double circle denote the results with $\Theta/\pi=0.05$, 0.15, 0.25, and 0.35.
}
\label{fig:dF-M18R10nu700}
\end{figure}

Moreover, with the maximum, $F_{\rm ob,max}$, and the minimum, $F_{\rm ob,min}$, in the pulse profiles shown in Fig. \ref{fig:Fob-M18R10nu700}, we plot the ratio of the difference between $F_{\rm ob,max}$ and $F_{\rm ob,min}$ to the average of $F_{\rm ob,max}$ and $F_{\rm ob,min}$, i.e.,
\begin{equation}
  \frac{\Delta F_{\rm ob}}{\bar{F}_{\rm ob}} \equiv \frac{2(F_{\rm ob,max} - F_{\rm ob,min})}{F_{\rm ob,max} + F_{\rm ob,min}},
\end{equation}
as a function of $|\Theta-i|/\pi$ in Fig. \ref{fig:dF-M18R10nu700}. From this figure, one can observe that $\Delta F_{\rm ob}/\bar{F}_{\rm ob}$ increases with $\Theta$ and $i$, fixing the value of $|\Theta-i|$, while $\Delta F_{\rm ob}/\bar{F}_{\rm ob}$ decreases with $|\Theta-i|$, fixing $\Theta$ or $i$.

\begin{figure*}
\begin{center}
\includegraphics[scale=0.38]{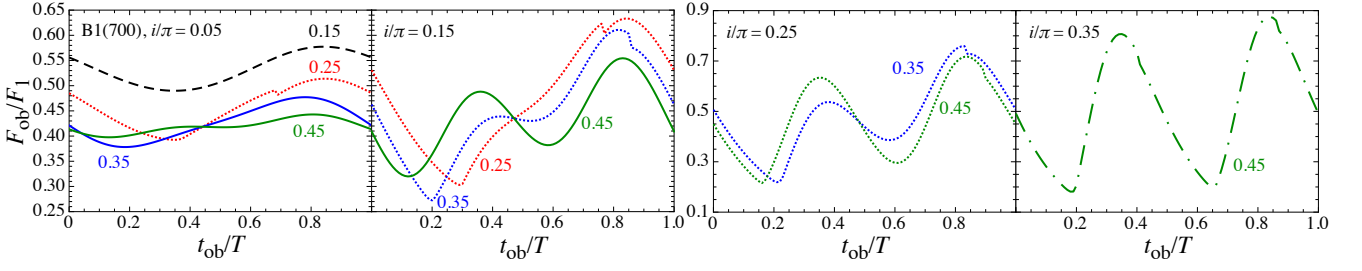}
\end{center}
\caption{
$F_{\rm ob}/F_1$ for B1(700) is shown as a function of $t/T$, where the  panels from left to right denote the results with $i/\pi=0.05$, 0.15, 0.25, and 0.35, while the number along with the line denotes the value of $\Theta/\pi$. The dashed, dotted, dot-dashed, and solid lines correspond to the light curves in the class I, II, III, and IV.
}
\label{fig:Fob-M18R13nu700}
\end{figure*}

\begin{figure*}
\begin{center}
\includegraphics[scale=0.5]{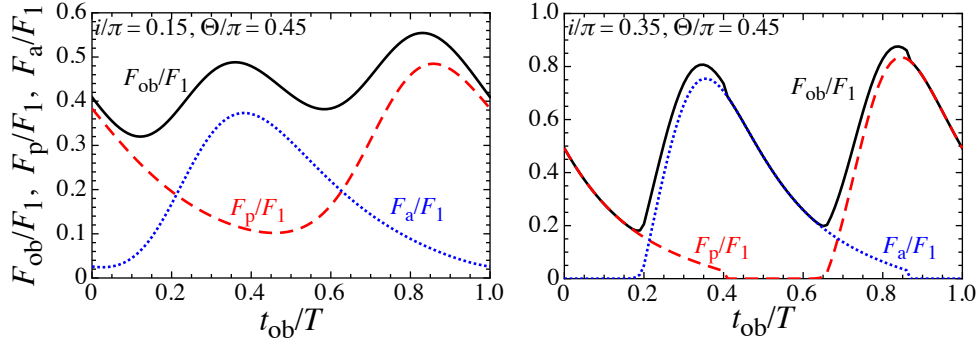}  
\end{center}
\caption{
For B1(700), the observed bolometric flux $F_{\rm ob}/F_1$, the flux from the primary spot $F_{\rm p}/F_1$, and the flux from the antipodal spot $F_{\rm a}/F_1$ are shown with the solid, dashed, and dotted lines with the angles of $i/\pi=0.15$ and $\Theta/\pi=0.45$ in the left panel and with the angles of $i/\pi=0.35$ and $\Theta/\pi=0.45$ in the right panel.
}
\label{fig:Fob-i035t045}
\end{figure*}

Next, the light curves for B1(700) with $i<\Theta$ are shown in Fig. \ref{fig:Fob-M18R13nu700}. As shown in the right panel of Fig. \ref{fig:class}, unlike the case of the model A, where $M/R_c=0.2658$, the classification how the hot spots are observed is not only the class IV but also I, II, or III for this stellar model ($M/R_c=0.2045$) with the adopted angles of $\Theta$ and $i$. To clarify such a classification, in Fig. \ref{fig:Fob-M18R13nu700} we show the light curves in the class I, II, III, and IV with the dashed, dotted, dot-dashed, and solid lines, respectively. From this figure, one can observe that the light curve in the class I or IV becomes smooth, because one or two hot spots can be observed in any time (see the left panel of Fig. \ref{fig:Fob-i035t045}). On the other hand, one also observes that the light curve in the class II or III has discontinuity, which comes from the fact that the hot spot sudden disappears when it enters into the invisible zone. In fact, one can observe a discontinuity in the class II light curve and two discontinuities in the class III light curve (see the right panel of Fig. \ref{fig:Fob-i035t045}).

\begin{figure*}
\begin{center}
\includegraphics[scale=0.38]{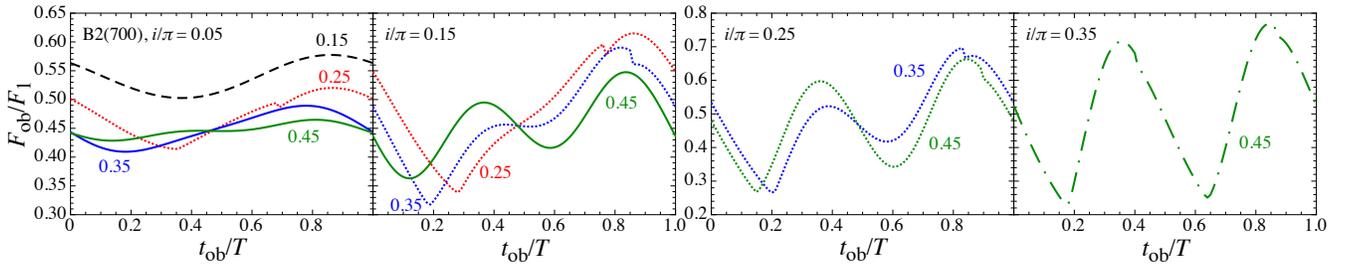}  
\end{center}
\caption{
Same as Fig. \ref{fig:Fob-M18R13nu700}, but for B2(700).
}
\label{fig:Fob-MR02045nu700}
\end{figure*}

In Fig. \ref{fig:Fob-MR02045nu700}, we show the light curve for the stellar model with the same compactness as in Fig. \ref{fig:Fob-M18R13nu700} but with different stellar radius. From this figure, one observes that the shape of light curves is very similar to that in Fig. \ref{fig:Fob-M18R13nu700}, but the amplitude is a little different from each other due to the effect of the special relativity. Considering that the light curve without the effect of the special relativity depends only on the stellar compactness, i.e., the light curve for a stellar model is the same as that for the stellar model with the same compactness even with different radius, the difference in the amplitude of the light curve due to the fast rotation may enable us to extract the information of the stellar radius.

\begin{figure}
\begin{center}
\begin{tabular}{cc}
\includegraphics[scale=0.5]{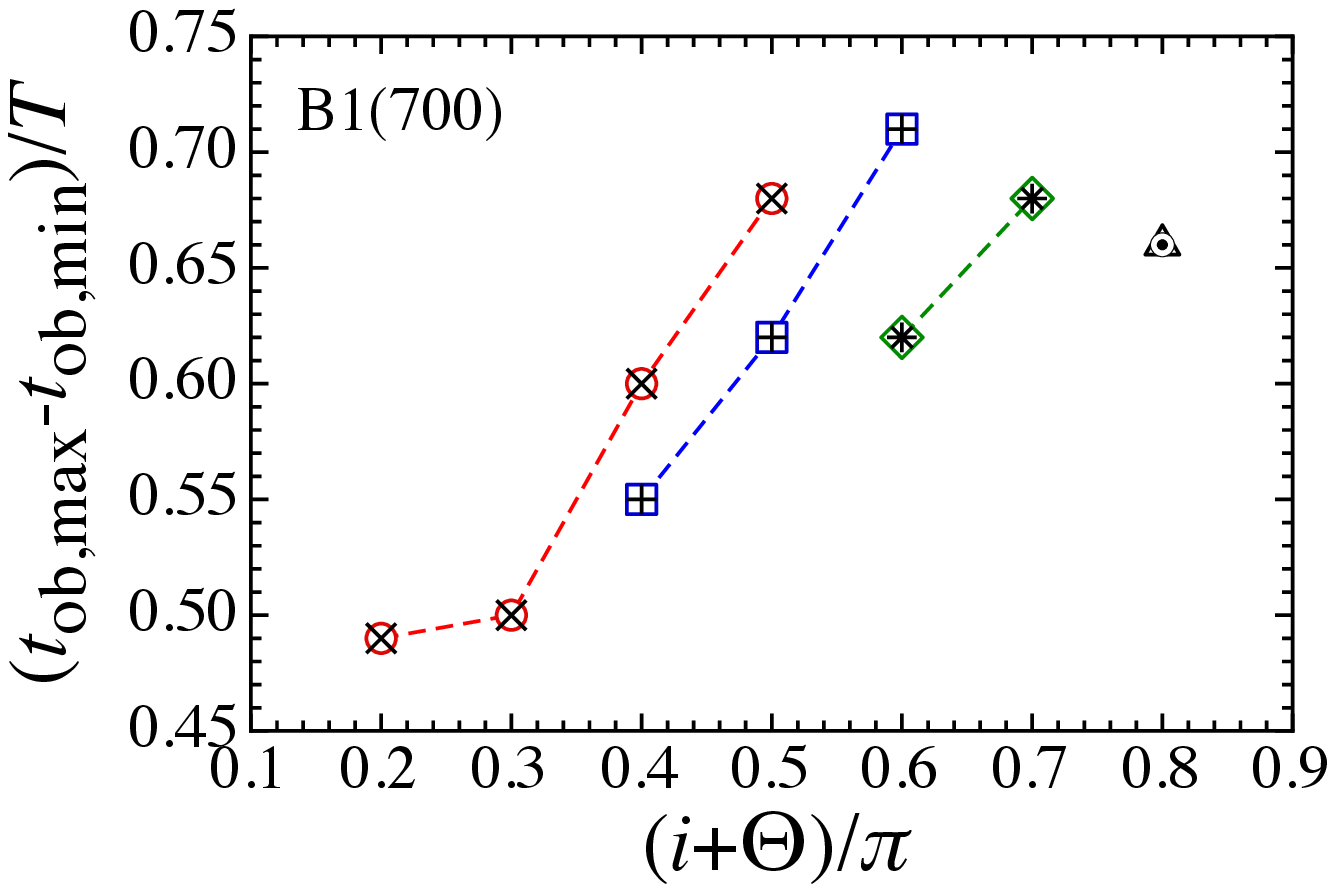}  &
\includegraphics[scale=0.5]{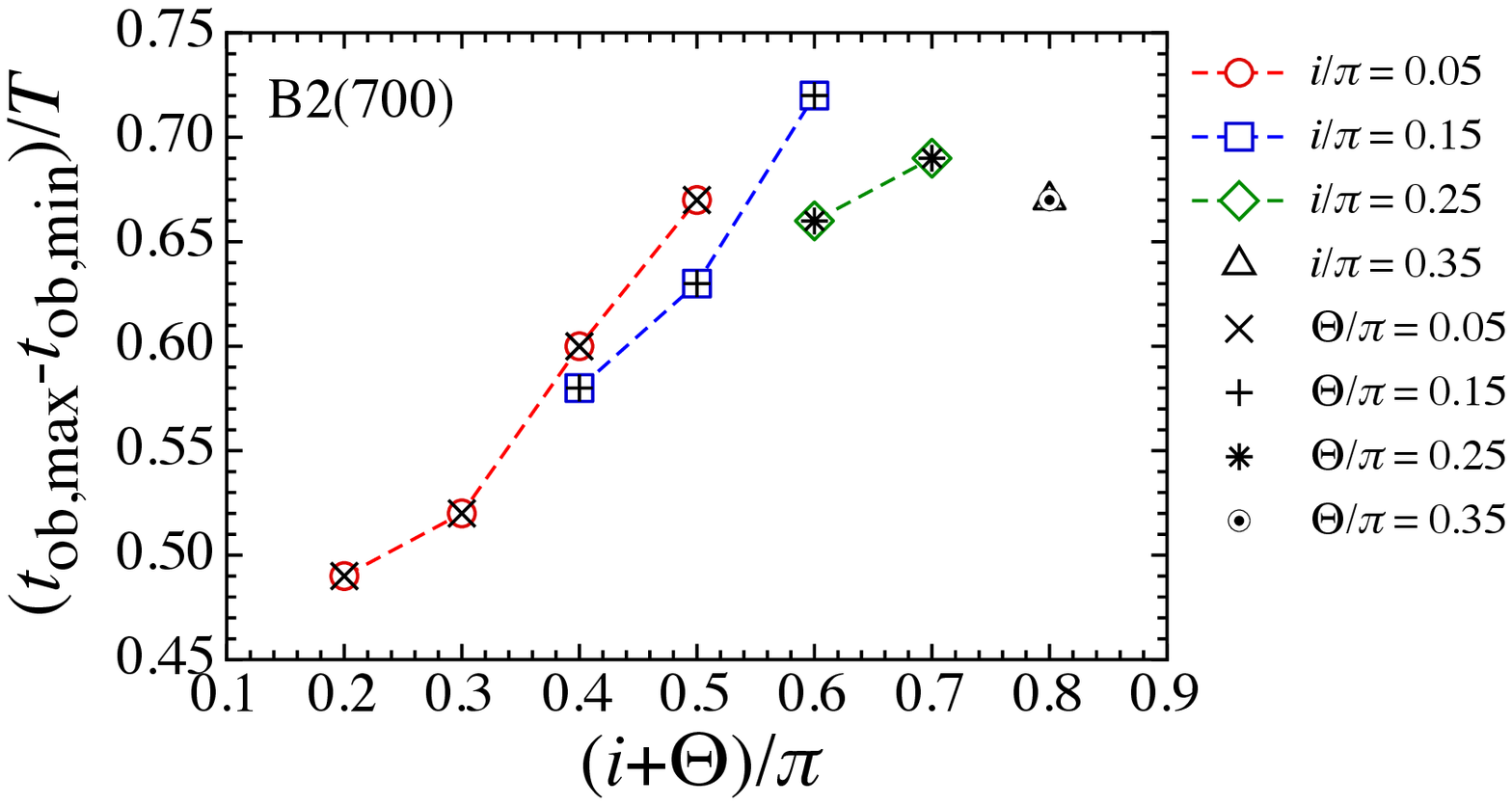}
\end{tabular}
\end{center}
\caption{
Same as Fig. \ref{fig:dtob-M18R10nu700}, but for B1(700) in the left panel and for B2(700) in the right panel.
}
\label{fig:dtob-M18R13nu700}
\end{figure}

\begin{figure}
\begin{center}
\begin{tabular}{cc}
\includegraphics[scale=0.5]{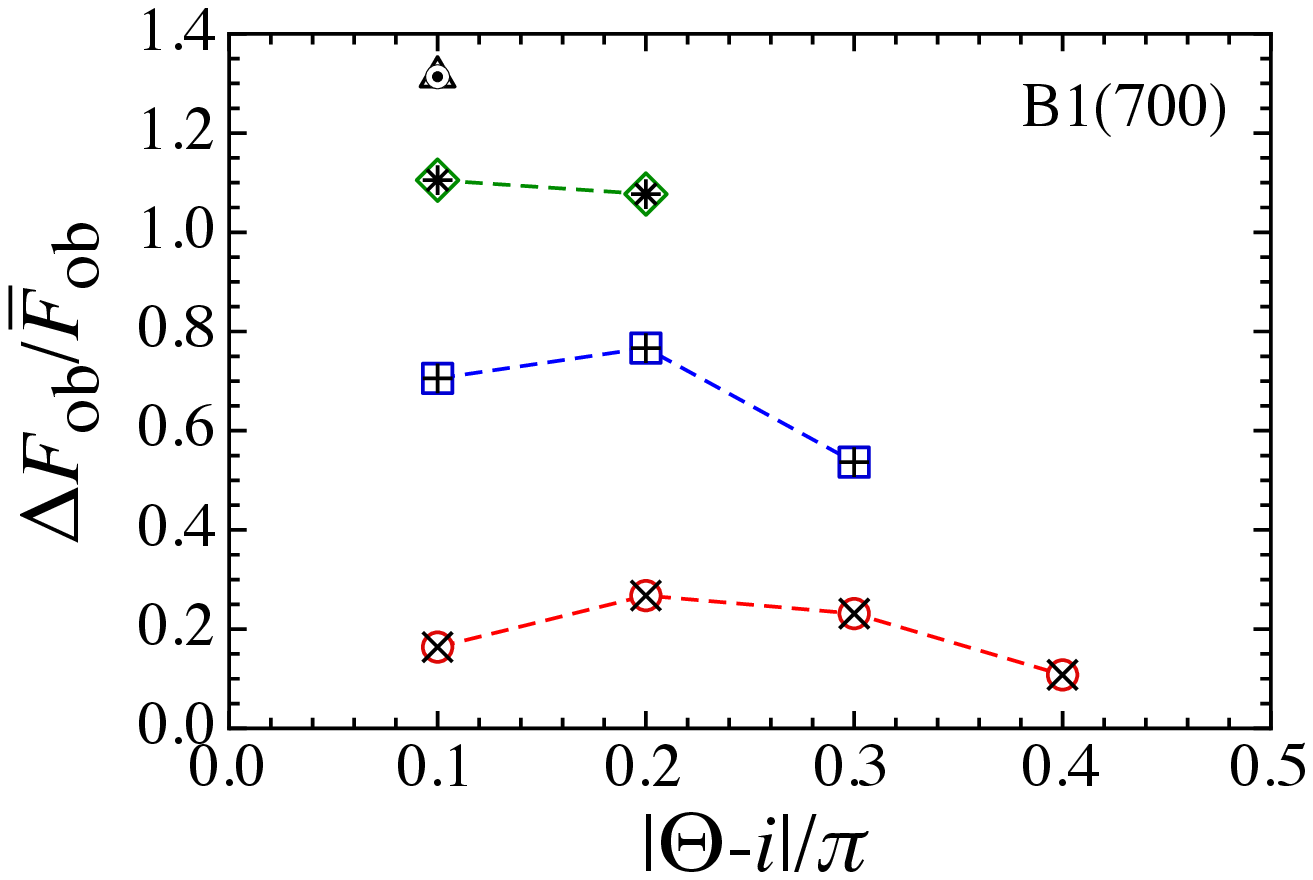}   &
\includegraphics[scale=0.5]{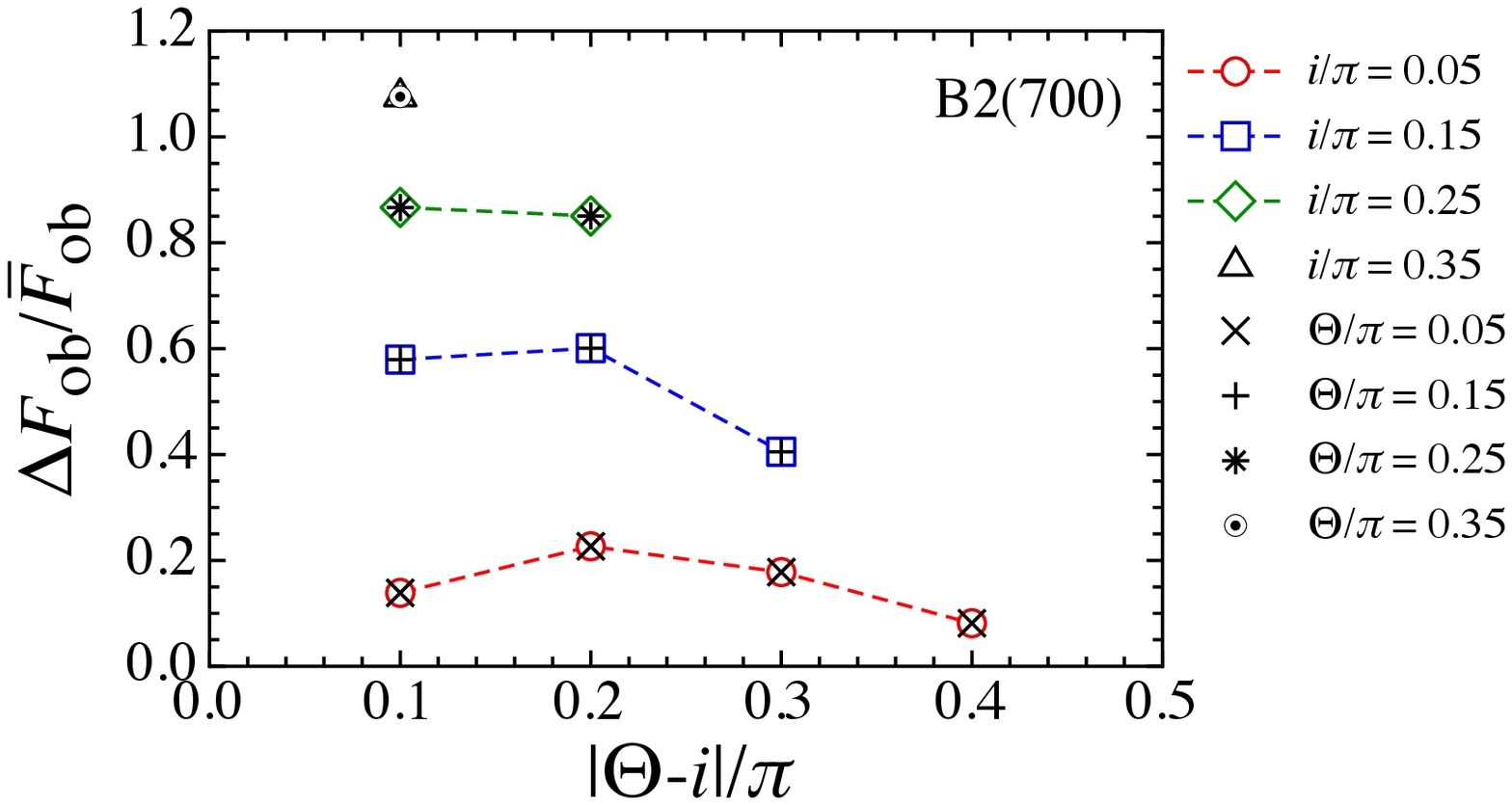} 
\end{tabular}
\end{center}
\caption{
Same as Fig. \ref{fig:dF-M18R10nu700}, but for B1(700) in the left panel and for B2(700) in the right panel.
}
\label{fig:dF-M18R13nu700}
\end{figure}

To see the difference in the light curves shown in Figs. \ref{fig:Fob-M18R13nu700} and \ref{fig:Fob-MR02045nu700}, as in Fig. \ref{fig:dtob-M18R10nu700}, the interval from the minimum up to the maximum in the light curve is shown in Fig. \ref{fig:dtob-M18R13nu700} as a function of $(i+\Theta)/\pi$, and as in Fig. \ref{fig:dF-M18R10nu700}, $\Delta F_{\rm ob}/\bar{F}_{\rm ob}$ is shown in Fig. \ref{fig:dF-M18R13nu700} as a function of $|\Theta-i|/\pi$, where in the both figures the left and right panels correspond to the results for B1(700) and B2(700), respectively. From Fig. \ref{fig:dtob-M18R13nu700} one can observe the difference even in the interval from the minimum up to the maximum in the light curve, because the time delay $\Delta t/T$ and Doppler factor $\delta$ depend on the stellar radius, even if the stellar compactness is constant. Meanwhile, from Fig. \ref{fig:dF-M18R13nu700}, we also confirm that the value of $\Delta F_{\rm ob}/\bar{F}_{\rm ob}$ becomes larger as the stellar radius is larger. Unlike the case for the model A ($M/R_c=0.2658$), since the light curves calculated for the model B ($M/R_c=0.2045$) are classified into various classifications how the hot spots can be observed, the dependence of $\Delta F_{\rm ob}/\bar{F}_{\rm ob}$ on $|\Theta-i|/\pi$ is not so simple, but still one can observe that $\Delta F_{\rm ob}/\bar{F}_{\rm ob}$ increases as $i$ or $\Theta$ increases, fixing the value of $|\Theta-i|$. In addition, we find that the value of $\Delta F_{\rm ob}/\bar{F}_{\rm ob}$ seems to become smaller as the stellar compactness increases, if the stellar radius is fixed, by comparing the results for A(700) in Fig. \ref{fig:dF-M18R10nu700} and for B2(700) in Fig. \ref{fig:dtob-M18R13nu700}.

\begin{figure}
\begin{center}
\begin{tabular}{c}
\includegraphics[scale=0.38]{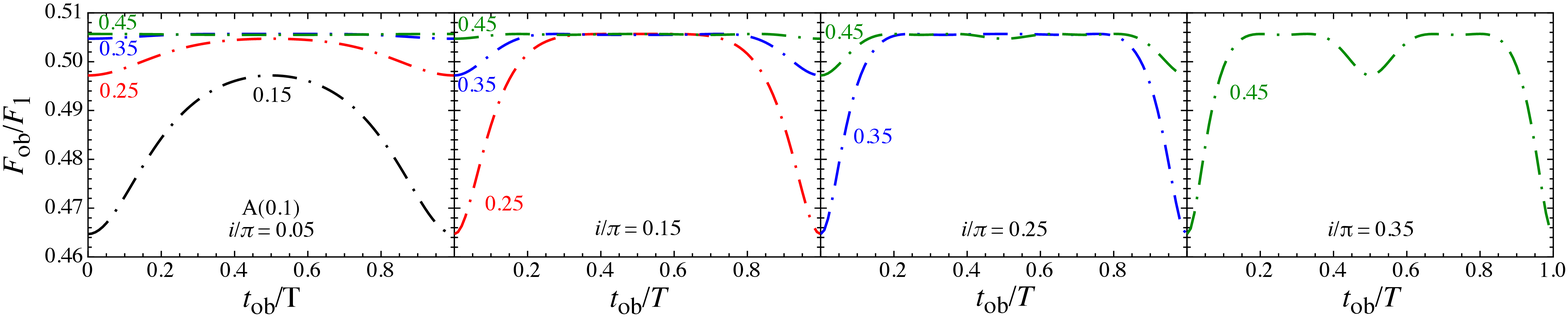}  \\
\includegraphics[scale=0.38]{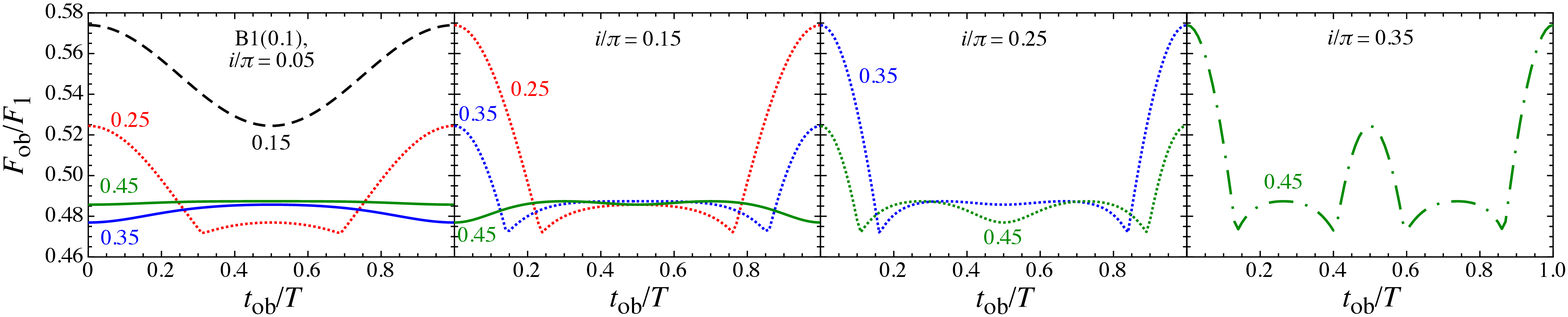} \\ 
\includegraphics[scale=0.38]{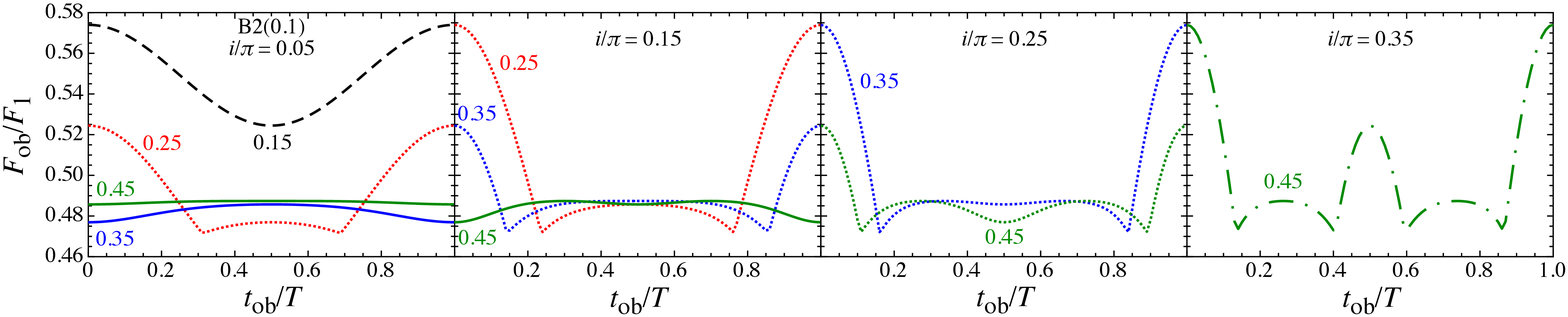}
\end{tabular}
\end{center}
\caption{
Pulse profiles are shown for A(0.1) in the upper panel, B1(0.1) in the middle panel, and B2(0.1) in the lower panel. In the same way as in Figs. \ref{fig:Fob-M18R13nu700} and \ref{fig:Fob-MR02045nu700}, the dashed, dotted, dot-dashed, and solid lines correspond to the light curves in the class I, II, III, and IV.
}
\label{fig:dF-Fobnu01}
\end{figure}

Finally, to compare the results for the fast rotating neutron stars with those for slowly rotating models, we show the light curves for the stellar model with $\nu=0.1$ Hz in Fig. \ref{fig:dF-Fobnu01}, where the upper, middle, and lower panels correspond to the results for the model A, B1, and B2, respectively. In this case, the special relativistic effect is almost negligible and the light curve at $t_{\rm ob}/T$ for $0\le t_{\rm ob}/T\le 0.5$ is almost the same as that at $1-t_{\rm ob}/T$ for $0.5\le t_{\rm ob}/T\le 1$. Additionally, we see that the light curve for the model B1 is the same as that for the model B2, i.e., the light curve depends only on the stellar compactness. Anyway, it is observed that the shape of the light curve shown in Fig. \ref{fig:dF-Fobnu01} is completely different from that for the fast rotating models.

\section{Conclusion}
\label{sec:V}

We systematically examine the light curve from the rotating neutron stars, taking into account the special relativistic effect as well as the time delay depending on the position of hot spots. For this purpose, we first derive the basic equations describing the bolometric flux from the spots with the general expression of the metric for the static, spherically symmetric spacetime, adopting the pointlike spot approximation as the previous studies. While the light curve from a slowly rotating neutron star depends only on the stellar compactness, we find that the light curve from a fast rotating neutron star depends not only on the stellar compactness but also the stellar radius. In fact, the amplitude of light curve becomes larger as the stellar radius increases and as the stellar compactness decreases. In addition, we find that, unlike the case for a slowly rotating neutron star, the pulse profiles become asymmetry under the interchange of the angle between the magnetic and rotational axes and the inclination angle. Even so, it is also found that the shape of light curves are the same as each other. From these results, one would determine the stellar compactness together with the stellar radius via careful observations of the light curve from the neutron star, using the spin frequency determined observationally. The information on the stellar compactness and radius would enable us to constrain the EOS for nuclear matter in a high density region.

\acknowledgments
This work was supported in part by Grant-in-Aid for Scientific Research (C) through Grant No. 17K05458 (H.S.) and No. 18K03652 (U.M.) provided by JSPS.

\appendix
\section{Similarity of the light curve}
\label{sec:a1}

In this appendix, we show that the shape of the light curve with $(i,\Theta)=(a,b)$ is the same as that with $(i,\Theta)=(b,a)$. Now, we write down the observed bolometric flux with specific angles of $(i,\Theta)$ as $F_{\rm ob}(i,\Theta)$. In the similar way, the time delay and Doppler factor with specific angles of $(i,\Theta)$ are also written as $\Delta t(i,\Theta)$ and $\delta(i,\Theta)$, respectively. As mentioned in the text, $F_{\rm ob}(i,\Theta)$ in the limit of $\delta\to 1$ and $\Delta t(i,\Theta)$ are symmetric under the interchange of $i$ and $\Theta$. Thus, one finds that $F_{\rm ob}(a,b)/F_{\rm ob}(b,a)=[\delta(a,b)/\delta(b,a)]^5$. Taking into account that $|\bm{v}|\cos\xi=-{\cal A}\sqrt{C(R)/A(R)}\omega\sin\Theta\sin i\sin(\omega t)$ and that ${\cal A}$ is symmetric under the interchange of $i$ and $\Theta$, the asymmetric part in $\delta(i,\Theta)$ under  the interchange of $i$ and $\Theta$ is just $1/\gamma(i,\Theta)=(1-|\bm{v}(i,\Theta)|^2)^{1/2}$, where $\gamma(i,\Theta)$ and $\bm{v}(i,\Theta)$ are the corresponding variables with specific values of $i$ and $\Theta$. Thus, one can show that
\begin{equation}
  \frac{F_{\rm ob}(a,b)}{F_{\rm ob}(b,a)} = \left[\frac{1-|\bm{v}(a,b)|^2}{1-|\bm{v}(b,a)|^2}\right]^{5/2},
\end{equation}
while the value of $\bm{v}(i,\Theta)$ is independent of the time. That is, $F_{\rm ob}(a,b)/F_{\rm ob}(b,a)$ is constant in time. This is a reason why the shape of the light curve with $(i,\Theta)=(a,b)$ is the same as that with $(i,\Theta)=(b,a)$.



\begin{thebibliography}{999}
\bibitem{highMR}
  If the stellar compactness is so high that the invisible zone disappears, i.e., $\psi_{\rm cri}>\pi$, the pulse profile can be qualitatively different from that with $\psi_{\rm cri}<\pi$ \cite{SM2018}. In fact, it happens when $M/R_c>0.284$ in the Schwarzschild spacetime. But, since such a situation is in a quite small region in the mass and radius plane as shown in Fig. 3 in Ref. \cite{SM2018}, we do not consider in this study.
\bibitem{TD-PB06}
   In Ref. \cite{PB06}, the time delay is defined as a difference from the travel time when $b=0$.
\end{thebibliography}
\end{document}